\documentclass[]{article}
\usepackage{graphicx}

\long\def\comment#1{}
\newcommand{\commentout}[1]{}

\newcommand{\secref}[1]{Section~\ref{#1}}
\newcommand{\eqref}[1]{Equation~\ref{#1}}
\newcommand{\figref}[1]{Figure~\ref{#1}}
\newcommand{\tabref}[1]{Table~\ref{#1}}

\newcommand{\denselist}{
      \setlength{\itemsep}{0pt}
      \setlength{\parsep}{1.5pt}
      \setlength{\topsep}{1.5pt}
      \setlength{\parskip}{2pt}
      \setlength{\partopsep}{0pt}
      \setlength{\labelwidth}{1em}
      \setlength{\labelsep}{0.5em} }

\newcommand{\bdesc}{\begin{description}\denselist}
\newcommand{\edesc}{\end{description}}

\begin{document}

\title{Social Information Processing in Social News Aggregation}
\author{Kristina Lerman\\
University of Southern California \\
Information Sciences Institute\\
4676 Admiralty Way\\
Marina del Rey, California 90292\\
lerman@isi.edu}

\maketitle

\begin{abstract}

The rise of the social media sites, such as blogs, wikis, Digg and
Flickr among others, underscores the transformation of the Web to
a participatory medium in which users are collaboratively creating,
evaluating and distributing information. The innovations introduced by social media
has lead to a new paradigm for interacting with information,
what we call 'social information processing'.
In this paper, we study how social news aggregator Digg exploits
social information processing to solve the problems of document recommendation and
rating. First, we show, by tracking stories over time, that social networks play an important role in
document recommendation.
The second contribution of this paper consists of two
mathematical models. The first model describes how collaborative rating and promotion of stories emerges from the independent
decisions made by many users. The second model describes how a user's influence,
the number of promoted stories and the user's social network, changes in time.
We find qualitative agreement between predictions of the model and user data gathered from Digg.

\end{abstract}

\section{Introduction}
The label {\em social media\/} has been attached to a quickly growing
number of Web sites whose content is primarily user driven. Examples
of such sites include the following: blogs (personal online journals
that allow users to share their thoughts and receive feedback on
them), Wikipedia (a collectively written and edited online
encyclopedia), and Flickr, Del.icio.us, and Digg (Web sites that allow
users to share, discuss, and rank photos, Web pages, and news stories
respectively). Other sites (e.g., Amazon's Mechanical Turk) allow
users to collaboratively find innovative solutions to hard problems.
The rise of social media underscores a transformation of the Web as
fundamental as its birth.  Rather than simply searching for, and
passively consuming, information, users are collaboratively creating,
evaluating, and distributing information.  In the near future,
new information-processing applications enabled by social media will
include tools for personalized information discovery, applications
that exploit the ``wisdom of crowds'' (e.g., emergent semantics and
collaborative information evaluation), deeper analysis of community
structure to identify trends and experts, and many other still difficult
to imagine.

Social media sites share four characteristics: (1)~Users create or
contribute content in a variety of media types; (2)~Users annotate
content with tags; (3)~Users evaluate content, actively by voting or
passively by using content; and (4)~Users create social networks by
designating other users with similar interests as contacts or friends.
We believe that social media facilitate new ways of
interacting with information.  In the process of using these sites,
users are adding rich metadata --- in the form of social networks,
annotations and ratings --- that enhances collaborative problem
solving through what we call \emph{social information processing}.

In this paper, we study how the social news aggregator Digg\footnote{http://digg.com} uses social information
processing to solve long standing problems, such as document
recommendation and rating.
The functionality of Digg is very simple:
users submit stories they find online, and other users rate these
stories by voting.  Digg also allows users to create social networks by adding other users
as friends and provides an interface to easily track their activities:
e.g., what stories users within their social network read and liked.   Each day, Digg promotes a handful of stories to
its front pages based on the stories' voting patterns.  Therefore, the
promotion mechanism does not depend on the decisions of a few editors,
but emerges from the activities of many users.  This type of
collective decision making can be extremely effective in breaking
news, often outperforming special-purpose algorithms.  For example,
the news of Rumsfeld's resignation in the wake of the 2006 U.S.\
Congressional elections broke Digg's front page within 3 minutes of
submission and 20 minutes before it was related by Google News
\cite{Rose2006Rumsfeld}. In addition to promoting news stories, Digg ranks users
by how successful they are at getting their stories promoted to the
front page.

Our first contribution is to show that social networks are used to discover new
interesting stories.  This type of \emph{social filtering} or
\emph{social recommendation} is an effective alternative to
collaborative filtering (CF), a popular recommendation technology used
by commercial giants like Amazon and Netflix. CF-based recommender
systems ask users to express their opinions by rating products, and
then suggest new products that were liked by other users with similar
opinions.  One noted problem with CF is that users are generally
resistant to rating products \cite{Konstan97grouplens}.  In contrast,
in social media sites, users express their tastes and preferences by
creating personal social networks. In the sites we studied, Flickr and
Digg, users generally take advantage of this feature, creating
networks of tens to hundreds (even thousands) of friends and using
them to filter content \cite{Lerman07digg,Lerman07flickr}.

Another outstanding problem in information processing is how to
evaluate the quality of documents.  This problem crops up daily in
information retrieval and Web search, where the goal is to find, among
the terabytes of data accessible online, the information that is most
relevant to a user's query.  The standard practice of search engines
is to identify all documents using the terms that appear in a user's
query, and rank the results according to their quality or importance.
Google revolutionized Web search by exploiting the link structure of
the Web, created through independent activities of many Web page
authors, in order to evaluate the contents of information on Web pages
\cite{PageRank}.  Similarly, social news aggregators Digg and
Reddit\footnote{http://reddit.com} evaluate the quality of news
stories using independent opinions of voters.

The second contribution of this paper consists of two
mathematical models that describe the dynamics of collaborative rating and the
evolution of users' rank.
We show
that the solutions to these models correctly predict the observed behavior of votes
received by actual stories and the behavior of users rank on Digg.

The paper is organized as follows.  In \secref{sec:digg}, we describe
Digg's functionality and features in greater detail.  In
\secref{sec:dynamics}, we study the dynamics of collaborative rating
of news stories on Digg.  We show in \secref{sec:filtering} that
social networks have a strong impact on the number of votes received
by a story through the mechanism of social filtering.  In
\secref{sec:mathvotes}, we develop a mathematical model of collective
voting and discuss its behavior.  We show how mathematical analysis
can be used to guide the design of collaborative voting systems.  In
\secref{sec:rank}, we develop a model of the dynamics of users' rank,
and compare its solutions to the observed changes in users' rank on
Digg. Finally, in \secref{limitations}, we discuss limitations of
mathematical modeling, and identify new directions for future
research.

\begin{figure*}[tbh]
  \center{\includegraphics[width=5in]{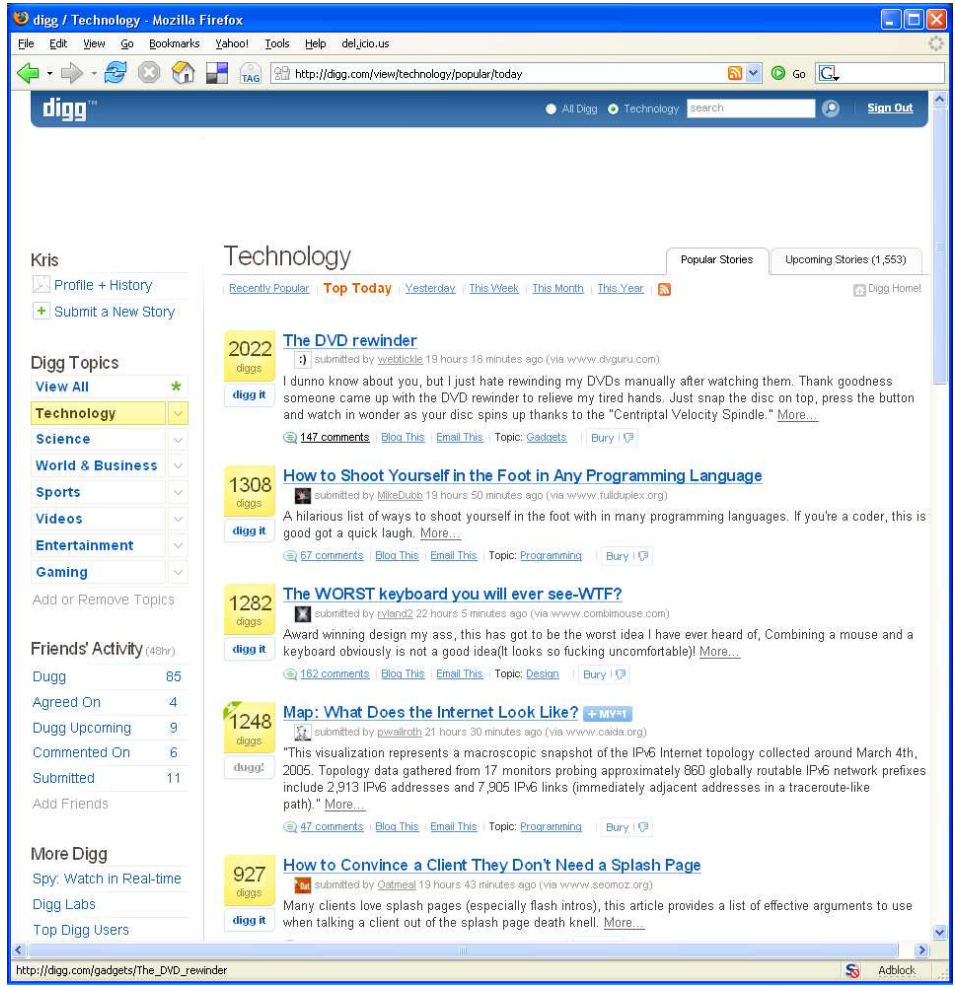}}
  \caption{Digg front page showing the technology
section}\label{fig:homepage}
\end{figure*}

\section{Anatomy of Digg} \label{sec:digg}

Digg is a social news aggregator that relies on users to submit
stories and moderate them. A typical Digg page is shown in
\figref{fig:homepage}. When a story is submitted, it goes to the
upcoming stories queue. There are 1-2 new submissions every minute.
They are displayed in reverse chronological order of being
submitted, 15 stories to a page, with the most recent story at the
top. The story's title is a link to the source, while clicking on
the number of diggs takes one to the page describing the story's
activity on digg: the discussion around it, the list of people who
voted on it, etc.

A user votes on a story by ``digging'' it. Digging a
story also saves it to user's history.  Digg also allows users to ``bury'' stories that are determined to be
spam, duplicates or contain inappropriate materials. ``Burying'' a story does not reduce its rating, as
voting a story down on the social news aggregator Reddit does. Rather,
if enough people have ``buried'' a story, it is permanently removed from Digg.

\subsection{Emergent front page}
When a story gets enough votes, it is promoted to the front page.
The vast majority  of people who visit Digg daily, or subscribe to
its RSS feeds, read only the front page stories; hence, getting to
the front page greatly increases the story's visibility. Although
the exact promotion mechanism is kept secret and changes
periodically, it appears to take into account the number of votes
the story receives. Digg's popularity is fueled in large part by
the phenomenon of the emergent front page which is formed by
consensus between many independent users.

% liken to Explore on Flickr
Other social media sites rely on similar mechanisms to showcase select content.
Every day the photo sharing site Flickr\footnote{http://flickr.com} chooses 500 most ``interesting'' of the newly
uploaded images to feature on the Explore page. The selection algorithm takes into
account how many times users view the image, comment on it or mark it as their favorite.\footnote{http://flickr.com/explore/interesting/}
Therefore, Flickr's ``interestingness'' algorithm also relies on emergent decision of many users.
Similarly, the social bookmarking site Delicious\footnote{http://del.icio.us} showcases the most popular
of the recently tagged Web pages.

\subsection{Social networks}
Digg allows users to designate others as friends and makes it easy
to track friends' activities. The Friends interface in the left column of the page in
\figref{fig:homepage} summarizes the number of stories friends have
submitted, commented on or liked recently. All these stories are
also are flagged with a green ribbon making them easy to spot.
Tracking activities of friends is a common feature of many
social media sites and is one of their major draws.
It offers a new paradigm for interacting with information
--- social filtering. Rather than actively searching for new
interesting content, or subscribing to a set of predefined topics,
users can put others to the task of finding and filtering
information for them.

\subsection{Top users} Until February 2007 Digg ranked users
according to how many of the user's stories were
promoted to the front page. User ranked number one had the highest number of front page stories.
If two users had an equal number of
front page stories, the one who was more active, commenting on and
digging more stories, received a higher rank. Clicking on the Top
Users link allowed one to browse through the ranked list of users.
There is speculation that ranking users increased competition,
leading some users to be more active in order to improve their
ranking. Digg discontinued making the list of top users publicly
available, citing concerns that marketers were paying top users to
promote their products and services \cite{WSJ}.

\section{Dynamics of collaborative rating} \label{sec:dynamics}

In order to study how the front page emerges from independent
decisions made by many users, we tracked both upcoming and front
page stories in Digg's technology section. We collected data by
scraping Digg site with the help of Web wrappers, created using
tools provided by Fetch Technologies\footnote{http://fetch.com/}:

\begin{description}
  \item[digg-frontpage] wrapper extracts a list of stories
from the first 14 front pages. For each story, it extracts
submitter's name, story title, time submitted, number of votes and
comments the story received, along with the list of the first 216
users who voted on the story.

  \item[digg-all] wrapper extracts a list of stories
from the first 20 pages in the upcoming stories queue. For each
story, it extracts the submitter's name, story title, time
submitted, number of votes and comments the story received.

  \item[top-users] wrapper extracts information about the top 1020 of
the recently active users.
  For each user, it extracts the number of stories
  that user has submitted, commented and voted on; number of stories
that have been promoted to
  the front page; number of profile views; time account was
  established; users's rank; the list of
  friends, as well as reverse friends or ``people who
  have befriended this user.''
\end{description}

\emph{Digg-frontpage} and \emph{digg-all} wrappers were executed
hourly over a period of a week in May 2006. \emph{Top-users} wrapper
was executed weekly starting in May 2006 to gather snapshots of the
social network of the top Digg users.

%\subsection{Time series of votes}
\begin{figure*}[tbh]
\begin{tabular}{cc}
  \includegraphics[height= 1.9in]{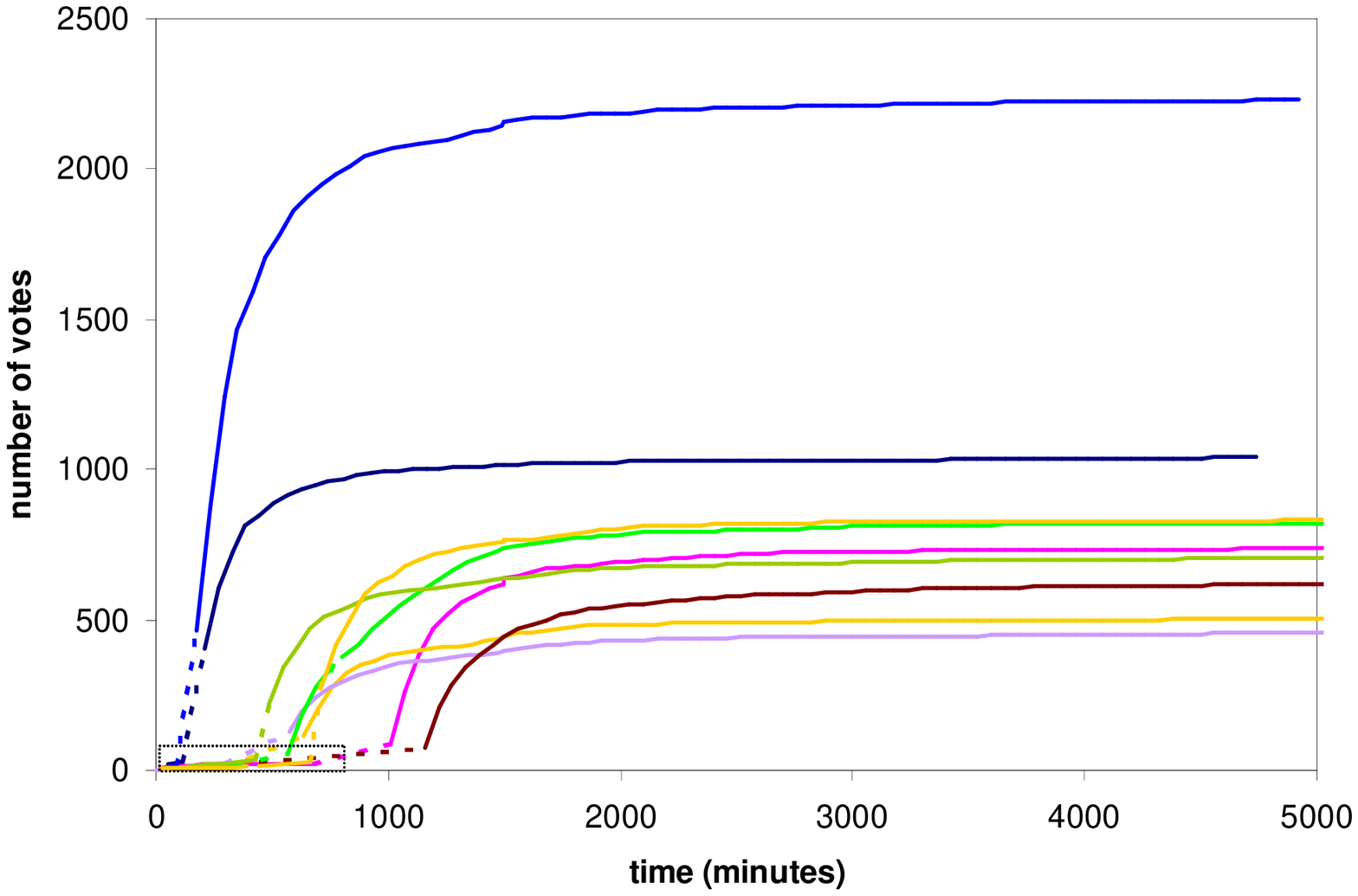} &
  \includegraphics[height= 1.9in]{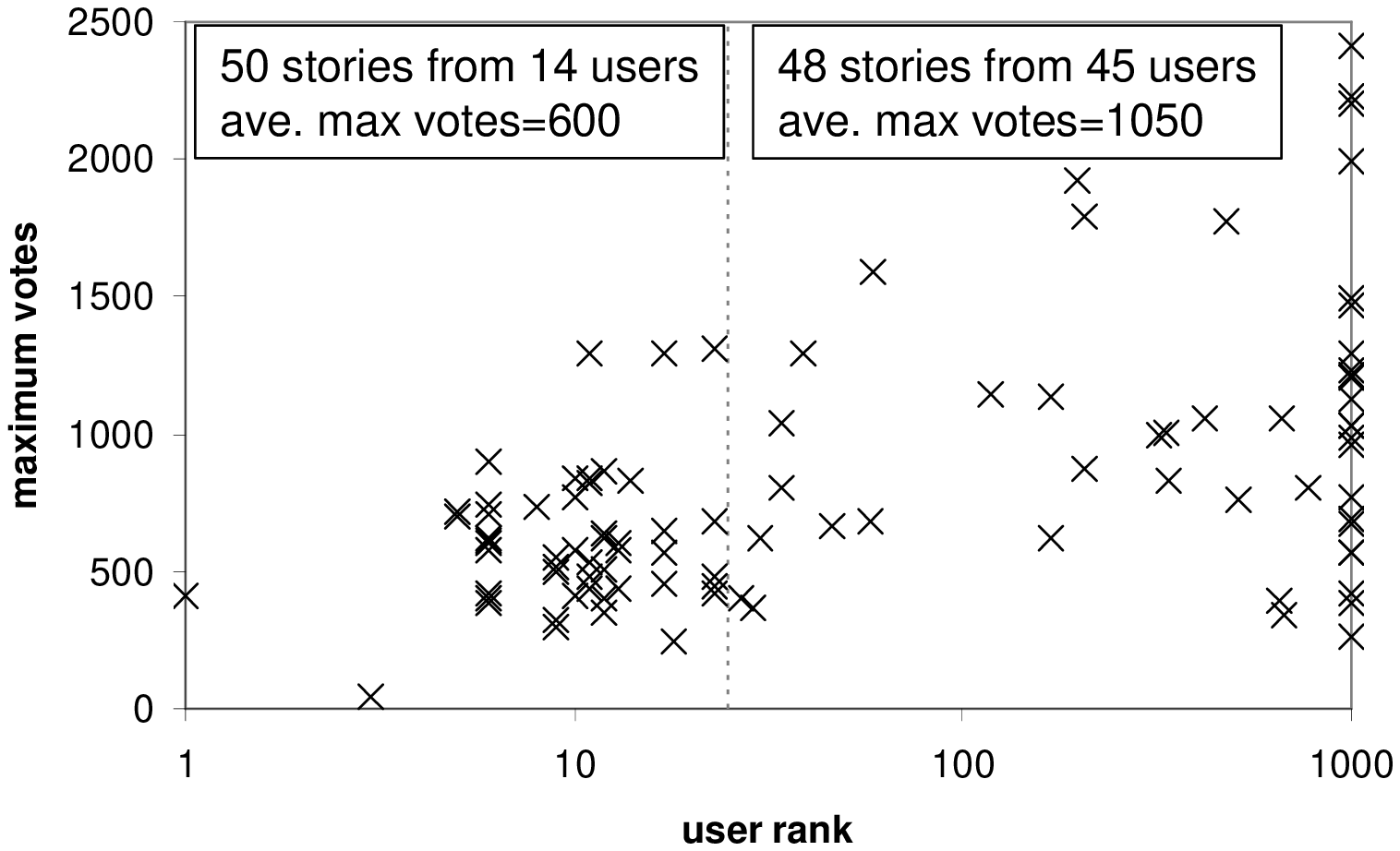} \\
  (a) & (b)
  \end{tabular}\caption{(a) Dynamics of votes of select stories over a period
of four days. The small rectangle in the lower corner highlights votes received
by stories while in the upcoming stories queue. Dashes indicate story's
transition to the front page.
%(b) Magnified view of the rectangle in (a). Dashed
%lines represent transition to the front page.
(b) Maximum number of votes received by stories during
the period of observation vs submitter's rank. Symbols on the right
axis correspond to low-rated users with rank$>1020$.} \label{fig:diggs}
\end{figure*}

We identified stories that were submitted to Digg over the course of
approximately one day and followed these stories over a period of
several days. Of the $2858$ stories that were submitted by $1570$
distinct users over this time period, only 98 stories by 60 users
made it to the front page. \figref{fig:diggs}(a) shows evolution of
the ratings (number of votes) of select stories. The basic dynamics of all stories appears the same. While in the upcoming stories queue, a story
accrues votes at some slow rate. Once it is promoted to the front page,
it accumulates votes at a much faster rate. As the story ages,
accumulation of new votes slows down, and the story's rating
saturates at some value. This value directly depends on how
interesting the story is to the Digg community.

It is worth noting that the top-ranked users are not submitting stories
that get the most votes. This is shown graphically in
\figref{fig:diggs}(b), which displays the maximum number of votes attained by stories in
the May dataset vs rank of the submitter. Slightly more than half of the stories came
from 14 top-rated users (rank$<25$) and 48 stories came from 45
low-rated users. The average ``interestingness'' of the stories
submitted by the top-rated users is $600$, almost half the average
``interestingness'' of the stories submitted by low-rated users. A
second observation is that top-rated users are responsible for
multiple front page stories. A look at the statistics about top
users provided by Digg shows that this is generally the case: of the
more than $15,000$ front page stories submitted by the top 1020
users, the top $3\%$ of the users are responsible for $35\%$ of the
stories.

\subsection{Social networks and social filtering}
\label{sec:filtering} If top-ranked users are not
submitting the most interesting stories, why are they so successful? We
believe that social filtering plays a role in helping promote stories to
the front page. As explained above, Digg's  allows users
to designate others as ``friends'' and offers an interface to easily track friends'
activities: the stories they have submitted, commented and voted on.
We believe that users employ the Friends interface to filter the tremendous
number of new submissions on Digg to find new interesting stories.

\begin{figure*}[tbh]
\begin{tabular}{cc}
  % after \\: \hline or \cline{col1-col2} \cline{col3-col4} ...
  \includegraphics[height=2.0in]{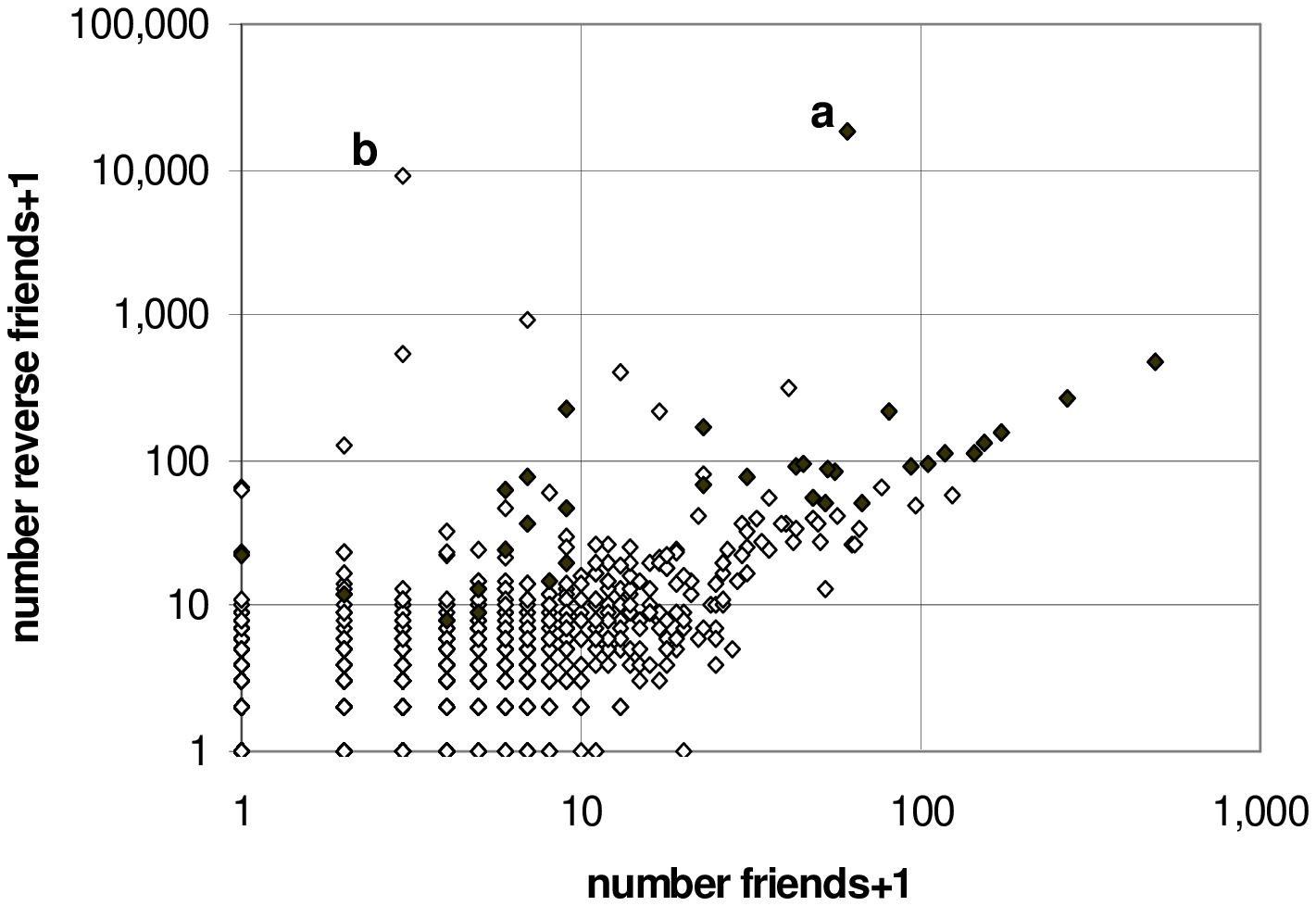}  &
  \includegraphics[height=2.0in]{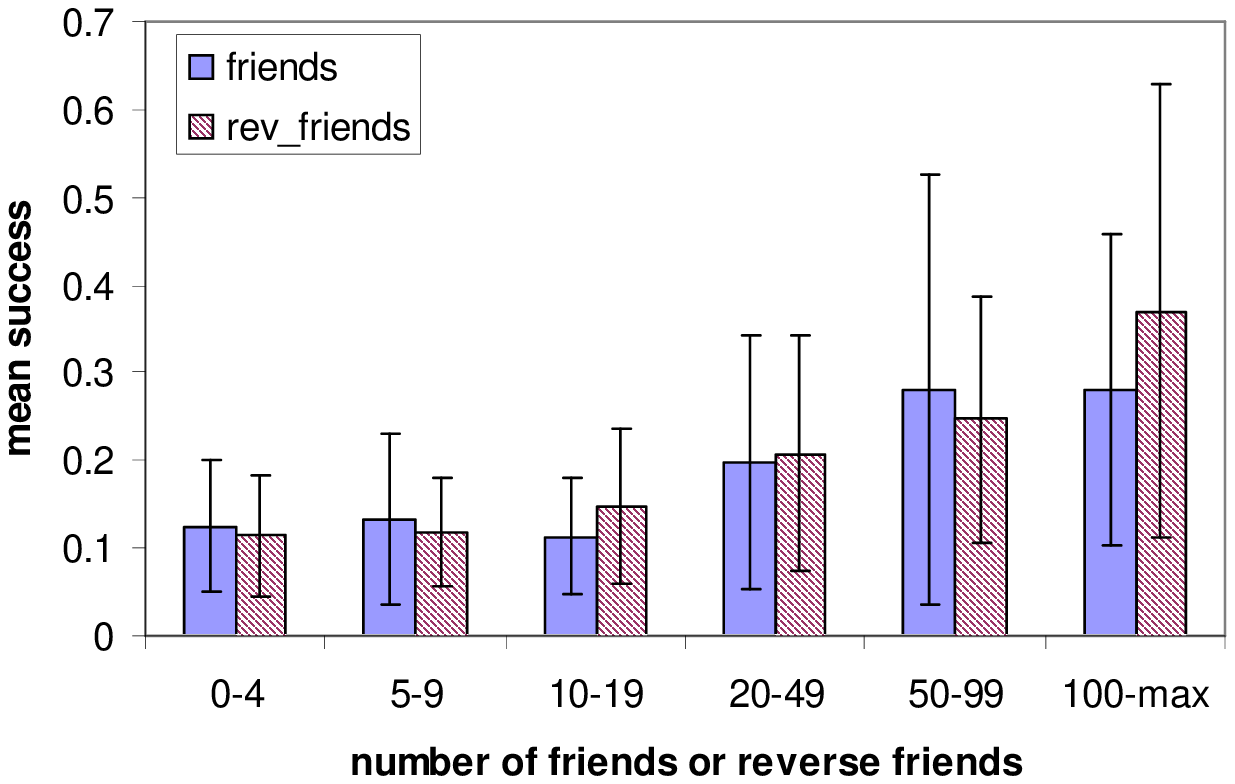}\\
  (a) & (b) \\
\end{tabular}
  \caption{(a) Scatter plot of the number of friends vs reverse friends for the top 1020 Digg users.
  (b) Strength of the linear correlation coefficient between user's success rate and the number
  of friends and reverse friends he has.
story}\label{fig:scatterplot}
\end{figure*}

Note that the friend relationship is asymmetric. When user $A$ lists user
$B$ as a \emph{friend}, user $A$ is able to watch the activities of
$B$ but not vice versa. We call $A$ the \emph{reverse friend} of
$B$. \figref{fig:scatterplot}(a) shows the scatter plot of
the number of friends vs reverse friends of the top 1020 Digg users
as of May 2006. Black symbols correspond to the top 33 users. For
the most part, users appear to take advantage of Digg's social
networking feature, with the top users having bigger social
networks. Users below the diagonal are watching more people than are
watching them (fans), while users above the diagonal are being
watched by more other users than they are watching (celebrities).
Two of the biggest celebrities are users marked $a$ and $b$ on
\figref{fig:scatterplot}(a). These users are $kevinrose$ and
$diggnation$, respectively, one of the founders of Digg and a
podcast of the popular Digg stories.

First, we present indirect evidence for social filtering on Digg by showing that
user's success is correlated with his social network size. A user's success
rate is defined as the fraction of the stories the user submitted
that have been promoted to the front page. We use the statistics
about the activities of the top 1020 users to show that users with
bigger social networks are more successful at getting their stories
promoted to the front page. In the analysis of the top 1020 user statistics, we only
include users who have submitted 50 or more stories (total of 514
users). The correlation between users's mean success rate and the size of their social
network is shown in \figref{fig:scatterplot}(b). Data was binned to improve statistics. Although the error
bars are large, there is a significant correlation between users's success rate and the size
of their social network, more importantly, the number of
reverse friends they have.

% show social filtering directly
In the sections below we present additional evidence that the Friends interface
is used to find new interesting stories. We show this by analyzing two
sub-claims: (a) \emph{users digg stories their friends submit}, and
(b) \emph{users digg stories their friends digg}. By ``digging'' the story, we mean that users like the story and
vote on it.

\subsubsection{Users digg stories their friends submit} In order to
show that users digg stories their friends submit, we used
 \emph{digg-frontpage} wrapper to collect
195 front page stories, each with a list of the first 216 users to vote
on the story ($15,742$ distinct users in total). The name of the submitter
is first on the list.

\begin{figure}[tbh]
 \center \includegraphics[width=4.0in]{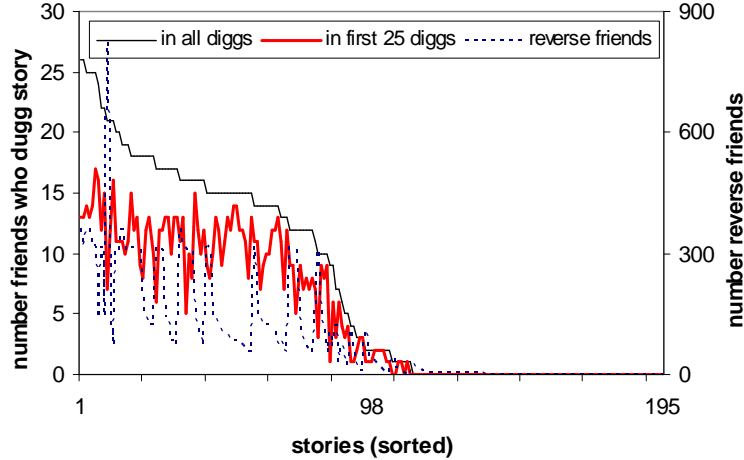}
  \caption{Number of voters who are also among the reverse friends of the user who submitted the
story}\label{fig:diggs-history}

\end{figure}

We can compare this list, or any portion
of it, with the list of the reverse friends of the submitter.
The thin line in \figref{fig:diggs-history} shows the number of voters
who are also among the reverse friends of the submitter, for all 195 stories. Dashed line shows the size of the social
network (number of reverse friends) of the submitter. More than half
of the stories (99) were submitted by users with more than 20
reverse friends, and the rest by poorly connected users.\footnote{These users
have rank $>1020$ and were not listed as friends of any of the 1020
users in our dataset. It is possible, though unlikely, that they
have reverse friends.}
All but two of the stories (submitted by users with
47 and 28 reverse friends) were dugg by submitter's reverse friends.

We use simple combinatorics \cite{Papoulis} to compute the probability that
$k$ of submitter's reverse friends could have voted on the story purely by
chance. The probability that after picking $n=215$ users randomly
from a pool of $N=15,742$ you end up with $k$ that came from a group
of size $K$ is $ P(k,n)={n\choose k} (p)^k (1-p)^{n-k}$, where
$p=K/N$. Using this formula, the probability (averaged over stories
dugg by at least one friend) that the observed numbers of reverse friends
voted on the story by chance is $P=0.005$, making it highly
unlikely.\footnote{If we include in the average the two stories that
were not dugg by any of the submitter's friends, we end up with a
higher, but still significant P=0.023.} Moreover, users digg stories
submitted by their friends very quickly. The heavy solid line in
\figref{fig:diggs-history} shows the number of reverse friends who
were among the first 25 voters. The probability that these numbers
could have been observed by chance is even less --- $P=0.003$. We
conclude that users digg --- or tend to like --- the stories their friends submit.
As a side effect, by enabling users to quickly digg stories
submitted by friends, social networks play an important role in
promoting stories to the front page.

\subsubsection{Users digg stories their friends digg}  Do social networks also help
users discover interesting stories that were submitted by unknown or poorly connected
users? Digg's Friends interface allows users to see the stories their friends have liked (voted on).
As well connected users digg stories, are others
within his or her social network more likely to read them?

\begin{figure*}
  % Requires \usepackage{graphicx}
\begin{tabular}{cc}
  % after \\: \hline or \cline{col1-col2} \cline{col3-col4} ...
  \includegraphics[height=2.2in]{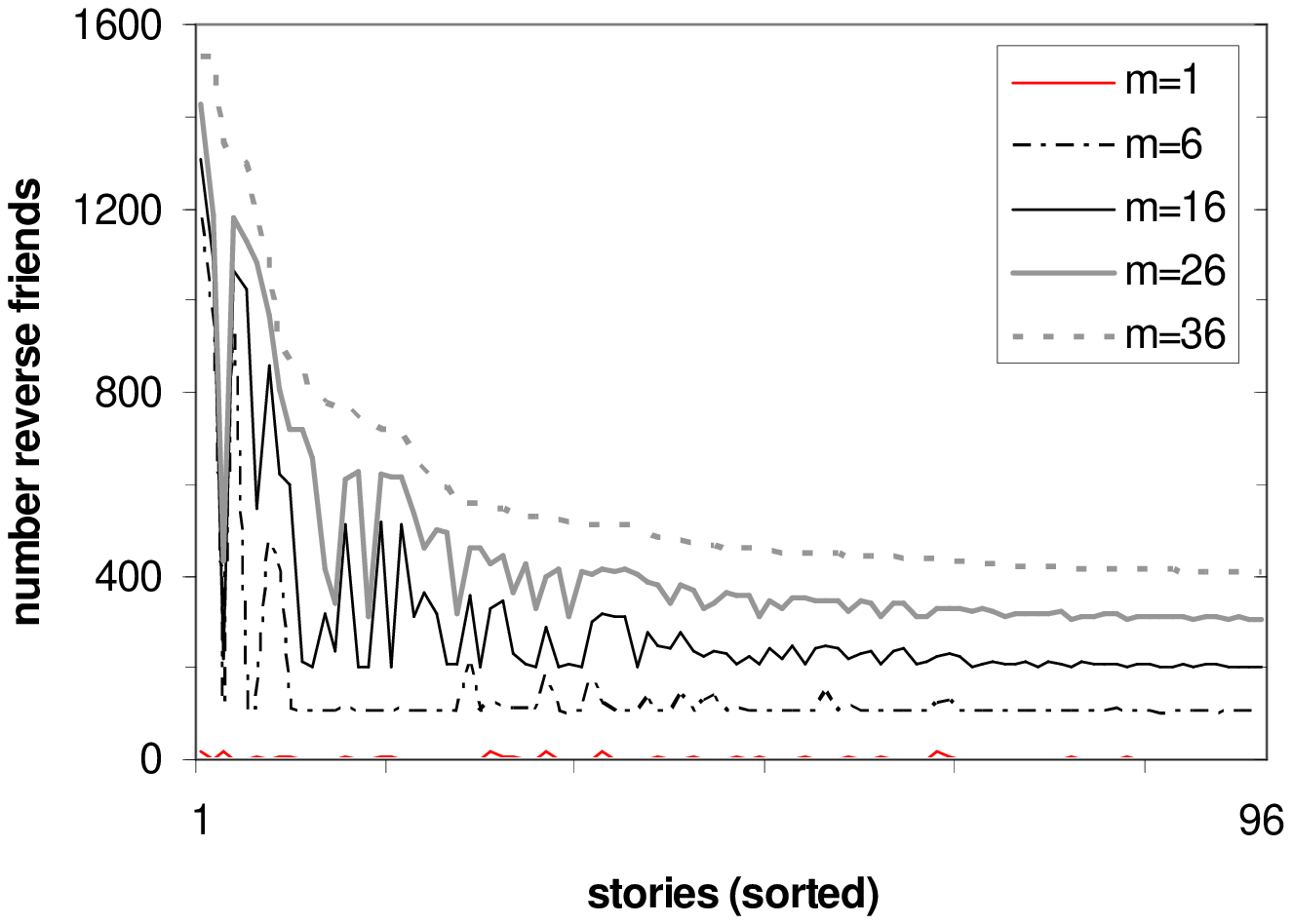}  &
  \includegraphics[height=2.2in]{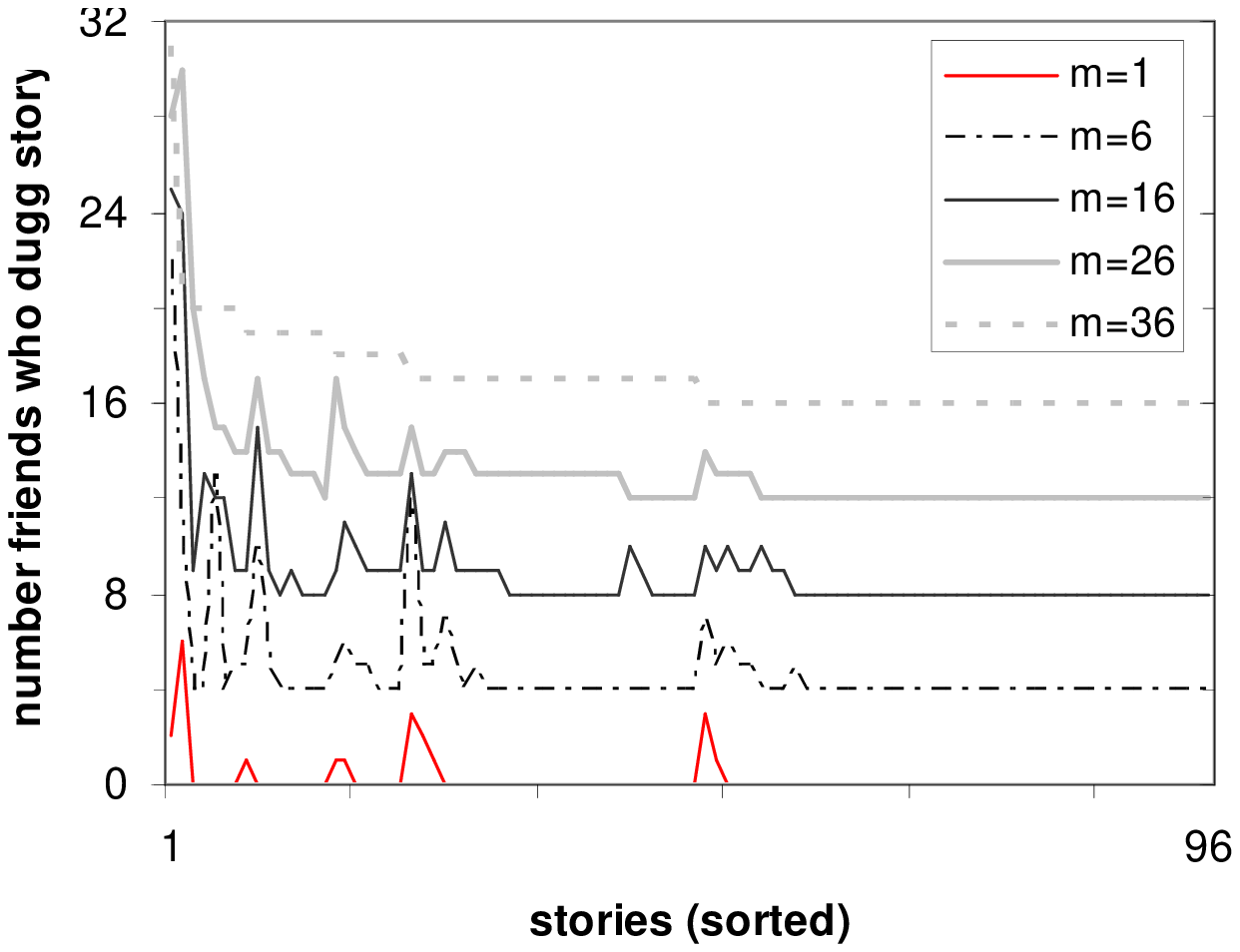}\\
  (a) & (b) \\
\end{tabular}
  \caption{(a) Number of reverse friends of the first $m$ voters for the stories submitted by unknown users.
  (b) Number of friends of the first $m$ voters who also voted on the stories.}\label{fig:history-unknown}
\end{figure*}

% dataset using week 5 user network data
% 99 stories, submitters had more than 20 friends. All but 2 of the stories were dugg by friends within the first 25 diggs.
% $P(m=1)=0.007$ within 25 diggs
% 96 stories were submitted by unknown users (fewer than 20 friends)
\begin{table*}
  \centering
\begin{tabular}{|ll|c|c|c|c|c|c|}
  \hline
  % after \\: \hline or \cline{col1-col2} \cline{col3-col4} ...
  & \textbf{diggers} & \textbf{m=1} & \textbf{m=6} & \textbf{m=16} & \textbf{m=26} & \textbf{m=36} & \textbf{m=46}
  \\ \hline
(a) & visible to friends & 34 & 75  & 94 & 96  & 96  & 96  \\
% dugg by friends (216) & 10 & 20 & 38 & 51 & 55 & 61 \\   \hline
 (b) & dugg by friends & 10 & 23 & 37 & 46 & 49 & 55 \\
  (c) & probability & 0.005 & 0.028 & 0.060 & 0.077 & 0.090 & 0.094 \\ \hline
\end{tabular}

  \caption{Number of stories posted by ``unknown'' users that were (a) made visible to other users through the
  digging activities of well-connected users, (b) dugg by friends of the first $m$ diggers within the next 25 diggs,
  and for the stories that were dugg by friends, (c) the average probability that the observed numbers of friends
  could have dugg the story by chance }\label{tbl:history-unknown}
\end{table*}

\figref{fig:history-unknown} shows how the activity of
well-connected users affected the 96 stories submitted by ``unknown'' users with fewer than 20 reverse friends.
$m=1$ corresponds to the user who submitted the story, while $m=6$
corresponds to the story's submitter and the first five users to
vote on it. Each line is shifted upward to aid visualization. Social networks appear to increase the story's
visibility. At the time of submission ($m=1$), only 26 of the 96
stories were visible to others within the submitter's social
network and ten of these were dugg by submitter's reverse friends within the first 25 votes.
After five more
users voted on the stories ($m=6$), 75 became visible to others through
the Friends interface, and of these 23 were dugg by friends.
After 25 users have voted on the stories, all 96 were visible through
the Friends interface, and almost half of these were dugg by
friends. \tabref{tbl:history-unknown} summarizes the observations
and presents the probability that the observed numbers of reverse friends
voted on the story purely by chance. The probabilities for $m=26$ through $m=46$ are
above the $0.05$ significance level, possibly reflecting the
increased visibility the story receives once it makes it to the
front page. Although the effect is not quite as dramatic as one in
the previous section, we believe that the data shows that users do
use the Friends interface to find new interesting stories that their friends have liked.

%\section{Tyranny of the minority and the promotion algorithm}
%\section{Design and consequences of collaborative rating systems}
\subsubsection{Changing the promotion algorithm}
\label{sec:promotion}
Digg's goal is to feature only the most interesting of the newly submitted stories on its
front page, and it employs aggregated opinion of thousands of its
users, rather than a few dedicated editors, to achieve this goal. Digg also allows users to create social networks by
designating others as friends and provides a seamless interface to
track friends' activities: what stories they submitted, liked, commented on, etc. By tracking stories
over time, we showed that social networks play an important role in
social filtering and recommendation.   Specifically, we showed (a)
users tend to like stories submitted by friends and (b) users tend
to like stories their friends read and like. Since some users are more active than others,
direct implementation of social filtering may lead to ``tyranny of
the minority,'' where a lion's share of front page stories come from
users with the most active social networks.  However,
precisely because these users are the most active ones, they play an
important role in filtering information and bringing to other
users's attention stories that would otherwise be buried in the
onslaught of new submissions.

\begin{figure}[tbh]
\center \includegraphics[width=3.0in]{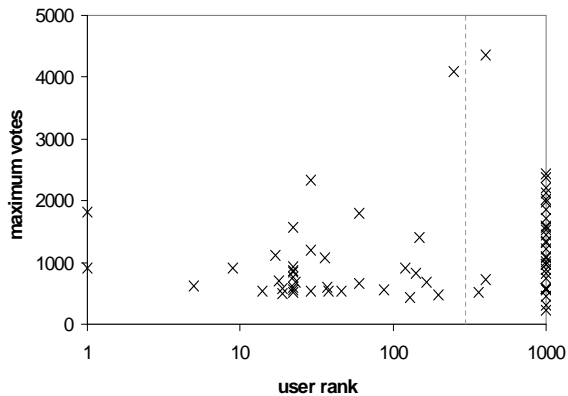}
  \caption{Maximum number of votes received by front page stories
  vs submitter's rank. Data was collected from stories submitted to Digg in early November 2006,
  after the change in the promotion algorithm. The vertical line divides the set in half.
  Symbols on the right hand axis correspond to low-rated users with rank$>1020$.
  }\label{fig:maxdiggs_new}
\end{figure}

Recently, a similar finding \cite{taylorhaywardblog} resulted in a
controversy on Digg \cite{USAToday}, in which users accused a
``cabal'' of top users of gaming the system by automatically voting on each other's stories.
The resulting uproar prompted Digg to change the algorithm it uses to promote stories. In
order to discourage what was seen as
``bloc voting,'' the new algorithm ``will look at the unique digging
diversity of the individuals digging the story'' \cite{diggblog}.
Analysis of the votes received by stories submitted in early November 2006
indicates that the algorithm change did achieve the desired effect of
reducing the top user dominance on the front page. Analysis of
the November data shows that of the 3072 stories submitted by 1865
users over a period of about a day, 71 stories by 63 users were promoted to
the front page. \figref{fig:maxdiggs_new} shows the maximum number
of votes received by these stories over a six day period vs
submitter's rank. Compared to the May data shown in \figref{fig:diggs}(b), the front
page now has a greater diversity of users, with fewer users
responsible for multiple front page stories: $1.2$ stories per submitter compared to $1.6$
stories in the May dataset. Another observation is that the rank distribution is less skewed towards
top-ranked users than before: half of the stories came from users with rank$<300$, rather than rank$<25$
in the May dataset. In addition, there is a smaller spread in the the mean interestingness of
stories submitted by higher and lower ranked users: 960 vs 1270 in the November dataset
compared with 600 vs 1050 in the May dataset.\footnote{The overall increase in the maximum
number of votes received by stories could reflect the growth of the Digg user base.}

 Although these changes in front page content may be seen as a positive development, the
change in the story promotion algorithm may have some unintended
consequences: it may, for example, discourage users from joining
social networks because their votes will be discounted. Mathematical analysis, described in the sections
below, can be used as a tool to investigate the consequences of changes in the story promotion algorithm.
Rather than being a liability,
however, social networks can be used
to personalize and tailor information to individual users, and drive
the development of new \emph{social search algorithms}. For example, Digg can offer
personalized front pages to every user that are based on his or her
friends' submission and voting history.

\subsection{Mathematical model of collaborative
rating}\label{sec:mathvotes} In this section we present a
mathematical model that describes how the number of votes received by a story
changes in time. Our goal is not only to produce a model
that can explain --- and predict --- the dynamics of collective voting
on Digg, but one that can also be used as a tool to study the
emergent behavior of different collaborative rating algorithms.

We parameterize a story by its
\emph{interestingness} coefficient $r$, which gives the probability
that a story will receive a positive vote once seen. The number of votes a
story receives depends on its \emph{visibility}, which simply means
how many people can see and follow the link to the story. The following factors
contribute to a story's visibility:
\begin{itemize}
\item visibility on the front page

\item visibility in the upcoming stories queue

\item visibility through the Friends interface

\end{itemize}

\begin{figure*}[tbh]
\begin{tabular}{cc}
  \includegraphics[height=1.9in]{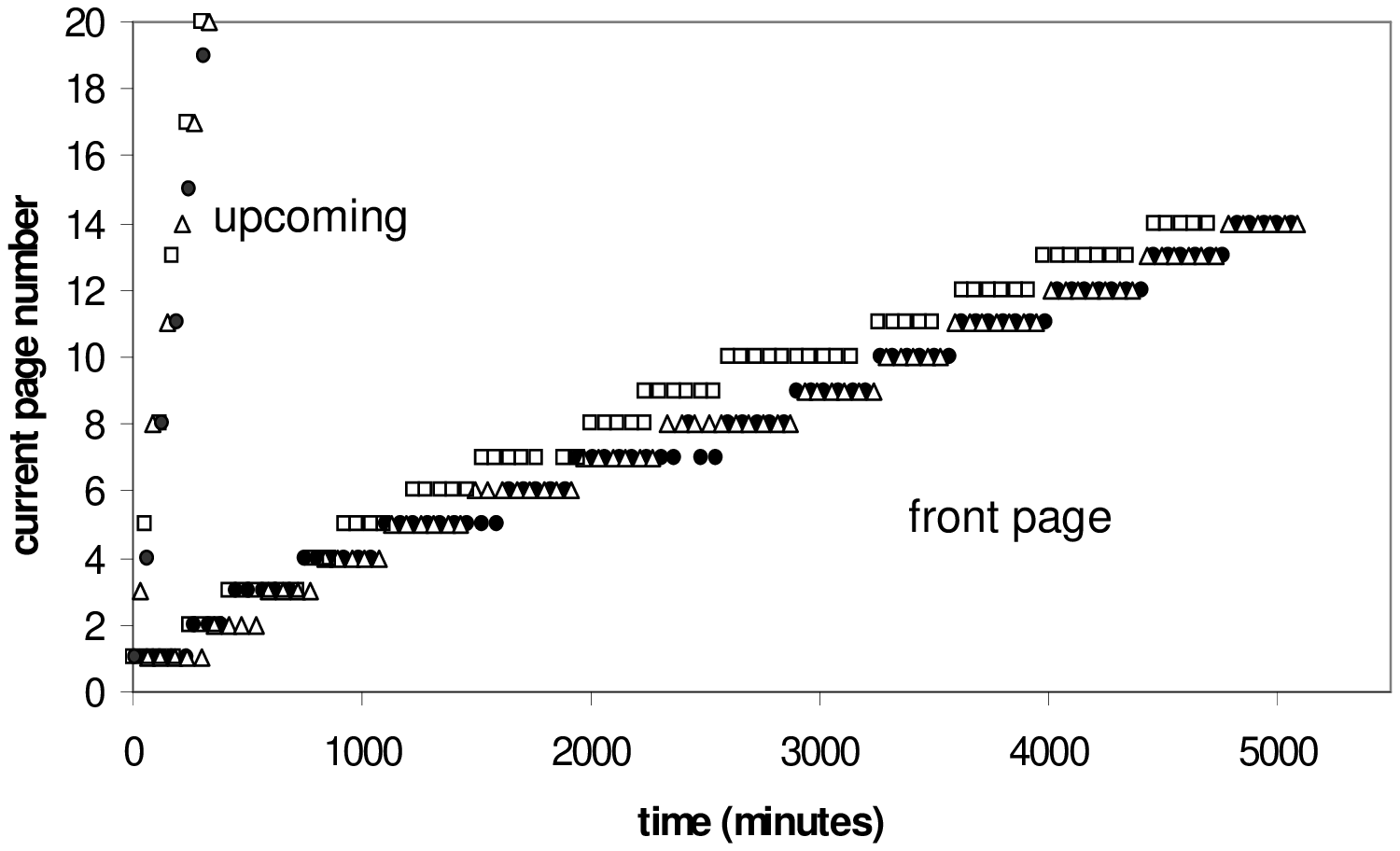} &
  \includegraphics[height=1.9in]{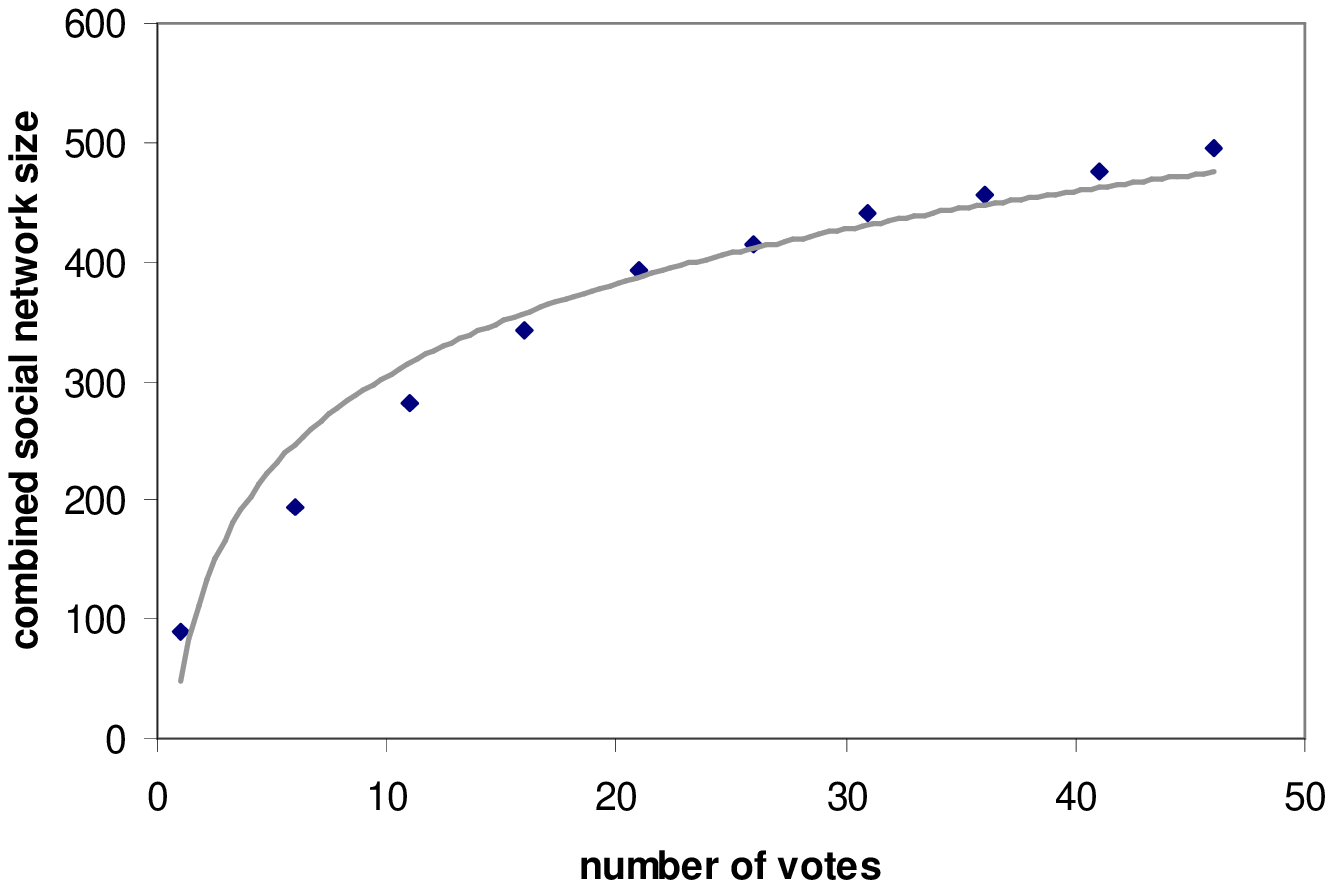} \\
  (a) & (b)
  \end{tabular}
\caption{(a) Current page number of a story on the upcoming stories
queue and the front page vs time for three different stories. (b)
Growth of the combined social network of the first 46 users to vote
on a story } \label{fig:params}
\end{figure*}

\subsubsection{Visibility on Digg's pages} A story's visibility on the
front page decreases as newly promoted stories push it farther down
the list. While we do not have data about Digg visitors' behavior,
specifically, how many proceed to page 2, 3 and so on, we propose to describe it by a
simple  model that holds that some fraction $c_f$ of the visitors to
the current front page proceed to the next front page. Thus, if $N$
users visit Digg's front page within some time interval, $c_f N$ users see the
second page stories, ${c}_f^2 N$ see the third page stories, and
${c}_f^{p-1} N$ users see page $p$ stories.

A similar model describes how a story's visibility in the upcoming
stories queue decreases as it is pushed farther down the list by the
newer submissions. If a fraction $c$ of Digg visitors proceed to the
upcoming stories section, and of these, a fraction $c_u$ proceed to
the next upcoming page, then $c c_u N$ of Digg visitors see second
page stories, and $c {c}_u^{q-1} N$ users see page $q$ stories.

\figref{fig:params}(a) shows how the current page number, both on
the upcoming stories queue and for the front page, changes
in time for three randomly chosen stories from the May data
set. The change in the story's current page number can be fit by
lines ${q,p}=k_{u,f}t$ with slopes $k_u=0.060$ pages$/$m ($3.60$
pages$/$hr) on the upcoming stories queue and $k_f= 0.003$ pages$/$m
($0.18$ pages$/$hr) on the front page.

\subsubsection{Visibility through the Friends interface}
The Friends interface offers the user ability to see the stories his
friends have (i) submitted, (ii) liked (voted on), (iii) commented
on during the preceding 48 hours or
 (iv) friends' stories that are still in the upcoming stories queue. Although it is
likely that users are taking advantage of all four features, we will
consider only the first two in the analysis. These features
closely approximate the functionality offered by other social media
sites: for example, Flickr allows users to see the latest images his
friends uploaded, as well as the images a friend liked (marked as
favorite). We believe that these features are more familiar to the
user and used more frequently than the other features.

\paragraph{Friends of the submitter}Let $S$ be the number of reverse friends the user who submits a
story has. As a reminder, these are the users who are watching the
submitter's activities. We assume that these users visit Digg daily,
and since they are likely to be geographically distributed across
many time zones, they see the submitted story at an hourly rate of
$a=S/24$. The story's visibility through the submitter's social
network is therefore $v_s=a\Theta (S-at) \Theta (48-t)$.
$\Theta(x)$ is a step function whose value is $1$ when $x \ge 0$ and
$0$ when $x<0$. The first step function accounts for the fact that
the pool of reverse friends is finite. As users from this pool read
the submitted story, the number of potential readers gets smaller.
The second function accounts for the fact that the story will be
visible through the Friends interface for $48$ hours after
submission only.

\paragraph{Friends of the voters} As the story is voted on, it becomes visible to more users through
the ``see the stories my friends liked'' part of the Friends
interface. \figref{fig:params}(b) shows the average size of $S_m$, the
combined social network of the first $m$ users to vote on the story.
Although $S_m$ is highly variable from story to story, it's
average value has consistent growth: $S_m=112.0*log(m)+47.0$.
Therefore, the story's visibility through the combined social
network of the first $m$ users who vote on it is $v_m=bS_m \Theta(h-m)
\Theta(48hrs-t)$, where $b$ is a scaling factor that depends on the length of the time interval:
for hourly counts, it is $b= 1/24$.

\subsubsection{Dynamics of voting}
In summary, the four factors that contribute to a story's visibility
are:
\begin{eqnarray}
% \nonumber to remove numbering (before each equation)
  v_f &=& {c}_f^{p(t)-1} N \Theta(m(t)-h) \\
  v_u &=& c {c}_u^{q(t)-1} N \Theta(h-m(t)) \Theta(24hrs-t)\\
  v_s &=&  a \Theta (S-at) \Theta (48hrs-t) \\
  v_m &=& b S_m \Theta(h-m(t)) \Theta(48hrs-t)
  \end{eqnarray}
\noindent $t$ is time since the story's submission.
We use a simple threshold to model how a story is promoted to the
front page. When the number of votes a story receives is fewer than
$h$, the story is visible in the upcoming queue; when $m \ge h$, the
story is visible on the front page. This seems to approximate Digg's
promotion algorithm as of May 2006, since in our dataset we did not see any front
page stories with fewer than 44 votes, nor did we see stories on the
upcoming queue with more than 42 votes. The second step function in
the $v_u$ term accounts for the fact that a story stays in the
upcoming queue for $24$ hours only, while step functions in $v_s$
and $v_m$ model the fact that it is visible in the Friends interface
for $48$ hours. The story's current page number on the upcoming
stories queue $q$ and the front page $p$ change in time according
to:
\begin{eqnarray}
  p(t) &=& (k_f (t-T_h)+1)\Theta(T_h-t) \\
  q(t) &=& k_u t + 1
\end{eqnarray}
\noindent with $k_u=0.060$ pages$/$min and $k_f=0.003$ pages$/$min.
$T_h$ is the time the story is promoted to the front page.

The change in the number of votes $m$ a story receives during a time
interval $\Delta t$ is
\begin{equation}\label{eq:diggs}
    \Delta m(t) =r (v_f + v_u + v_s + v_m )\Delta t
\end{equation}
\noindent where $r$ is the story's interestingness --- the
probability it will receive a vote when viewed.

\subsubsection{Solutions}

We solve Equation~\ref{eq:diggs} subject to the initial conditions
$m(t=0)=1$, $q(t=0)=1$, because a newly submitted story appears on
the top of the upcoming stories queue and it starts with a single
vote, that coming from the submitter himself. The initial condition
for the front page is $p(t<T_h)=0$, where $T_h$ is the time the
story was promoted to the front page. We take $\Delta t$ to be one
minute. The solutions of Equation~\ref{eq:diggs} show how the number of
votes received by a story changes in time for different values of
parameters $c$, $c_u$, $c_f$, $r$ and $S$. Of these, only the last
two parameters
--- the story's interestingness $r$ and the size of the submitter's
social network $S$ --- change from one submission to another.
Therefore, we fix values of the first three parameters $c= 0.3$,
$c_u=0.3$ and $c_f=0.3$ and study the effect $r$ and $S$ have on the
number of votes the story receives. We also fix the rate at which
visitors visit Digg at $N=10$ users per minute. The actual visiting
rate may be vastly different, but we can always adjust the other
parameters accordingly. We set the promotion threshold to $h=40$.

\begin{figure*}[tbh]
\begin{tabular}{cc}
$S=0$ & $S=80$ \\
  \includegraphics[height=2.2in]{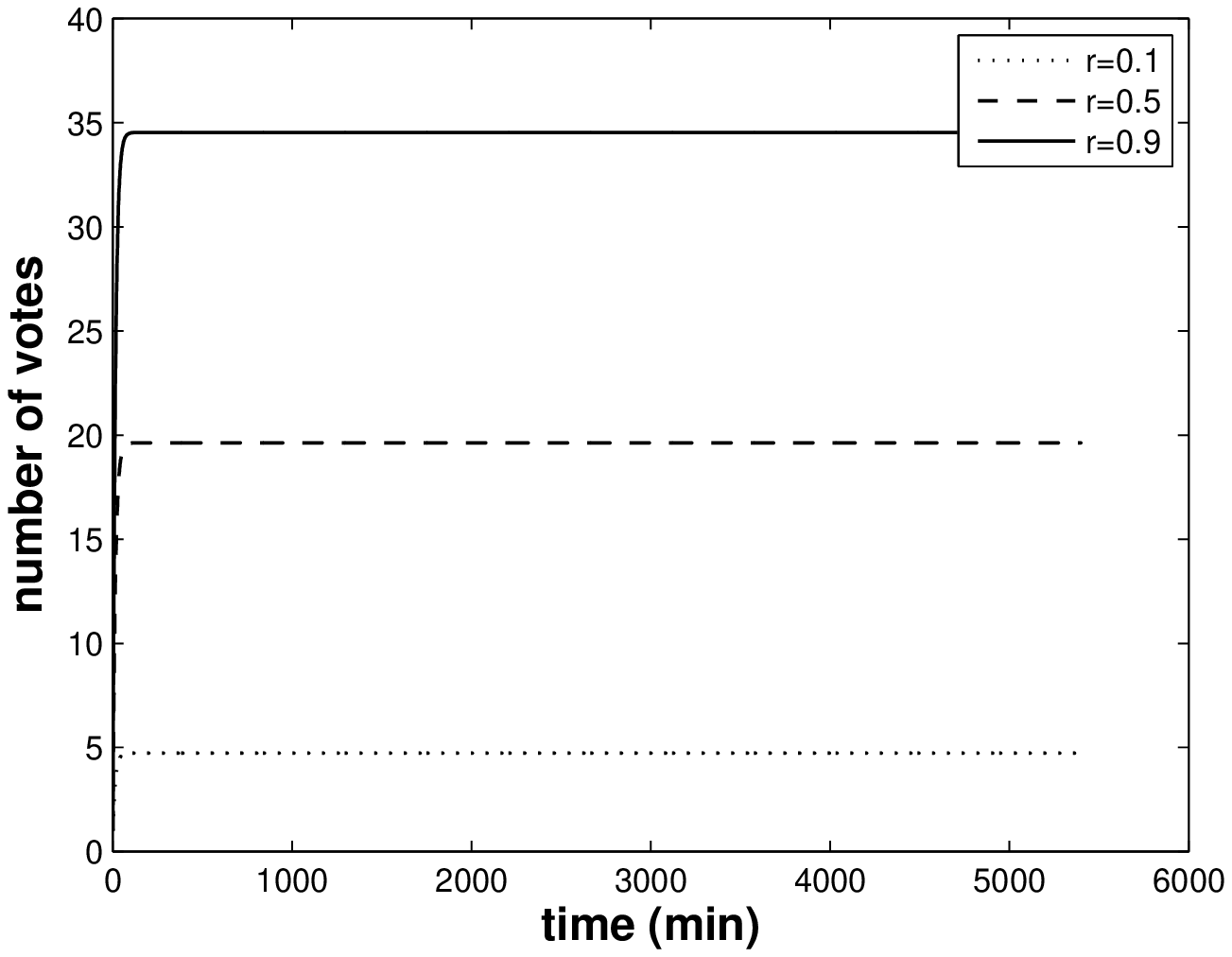} &
  \includegraphics[height=2.2in]{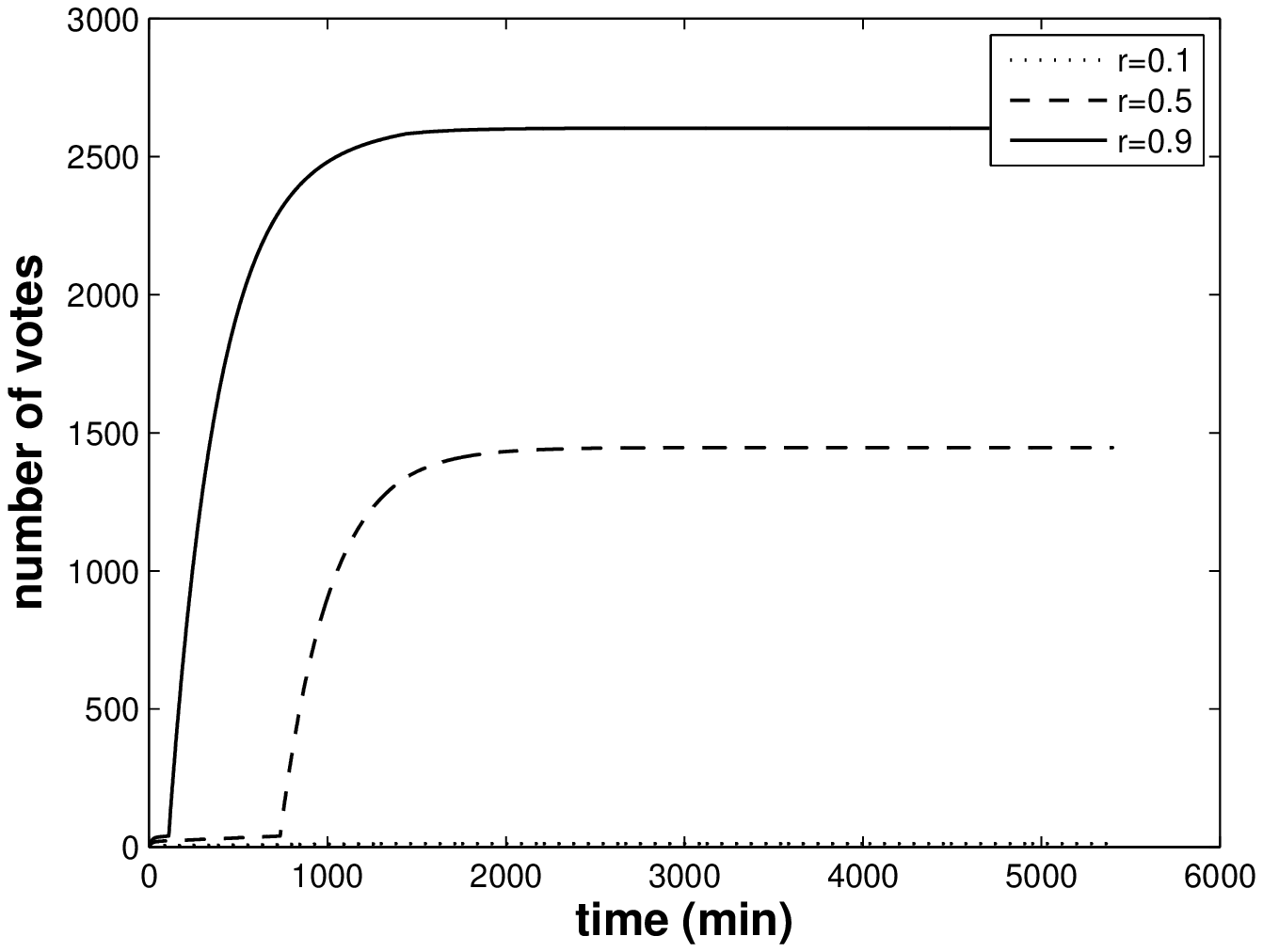}\\
  (a) & (b) \
  \end{tabular}
\caption{Effect of the submitter's social network on the evolution
of a story's rating. Votes received by a story posted by (a) an
unknown user with $S=0$ and a (b) connected user with $S=80$. }
\label{fig:votes-S}
\end{figure*}

First, we show that introducing social recommendation via the
Friends interface allows stories with smaller interestingness
parameter $r$ to be promoted to the front page. Suppose the Friends
interface only allows users to read the stories their friends
submit. \figref{fig:votes-S} shows how the ratings of three stories
with $r= 0.1$ , $r=0.5$ and $r=0.9$ change in time. For the chosen
parameter values, a story posted by an unknown user ($S=0$) never
gathers enough votes to exceed the promotion threshold $h$. Even a
highly interesting story with $r= 0.9$ languishes in the upcoming
stories queue until it eventually disappears. In fact, we can obtain
an analytic solution for the maximum number of votes a story can
receive on the upcoming stories queue, without the social network
effect being present. We set $v_f=v_s=v_m=0$ and convert
Equation~\ref{eq:diggs} to a differential form by taking $\Delta t
\rightarrow 0$:
\begin{equation}
\frac {dm}{dt}=r c c_u^{k_u t}N
\end{equation}
\noindent The solution of the above equation is
$m(T)=rcN(c_u^{k_uT}-1)/(k \log {c_u}) + 1$. Since $c_u<1$, the
exponential term will vanish for large times and leave us with
$m(T\rightarrow \infty)=-r c N/(k_u \log {c_u})+1 \approx 42 r +1$.
Hence, the maximum rating a story can receive on the upcoming pages
only is 43. Since the threshold on Digg appears to be set around
this value, no story can be promoted to the front page without other
effects, such as users reading stories through the Friends
interface.

A story posted by a user with $S=80$ will be promoted to the front
page if it is interesting enough, e.g., with $r=0.5$ and $r=0.9$ as
shown in \figref{fig:votes-S}(b). Note that the more interesting
story is promoted faster than a less interesting story
--- a general feature of collective voting. Stories posted by more
highly connected users  follow the same pattern, although the
interestingness value a story needs in order to get promoted is
decreased. For example, a story with $r=0.1$ posted by a user with
$S=400$ will be promoted to the front page.

\begin{figure*}[tbh]
\begin{tabular}{cc}
  empirical data & model prediction \\%& variable threshold \\
  \includegraphics[height=2.2in]{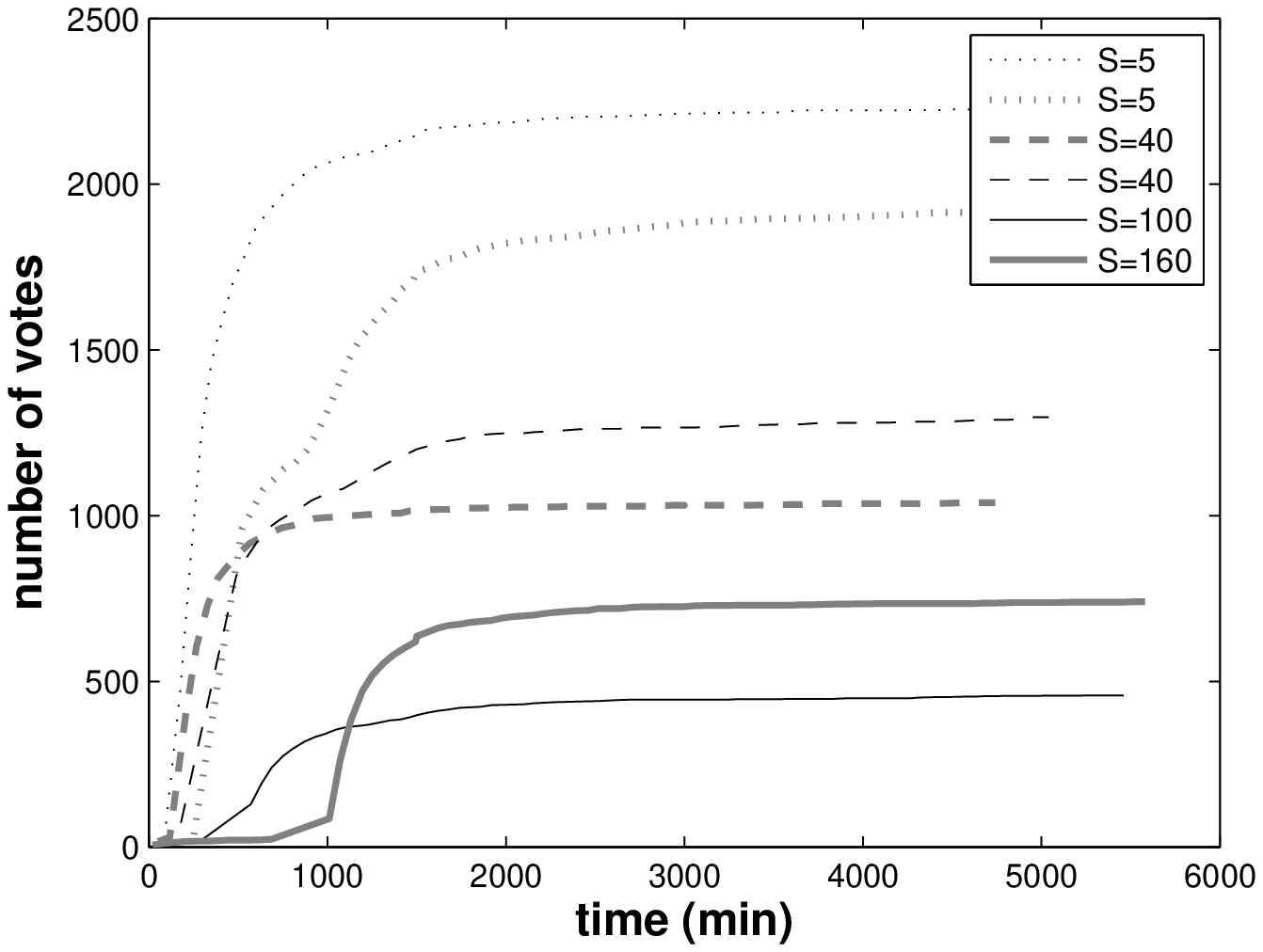} &
  \includegraphics[height=2.2in]{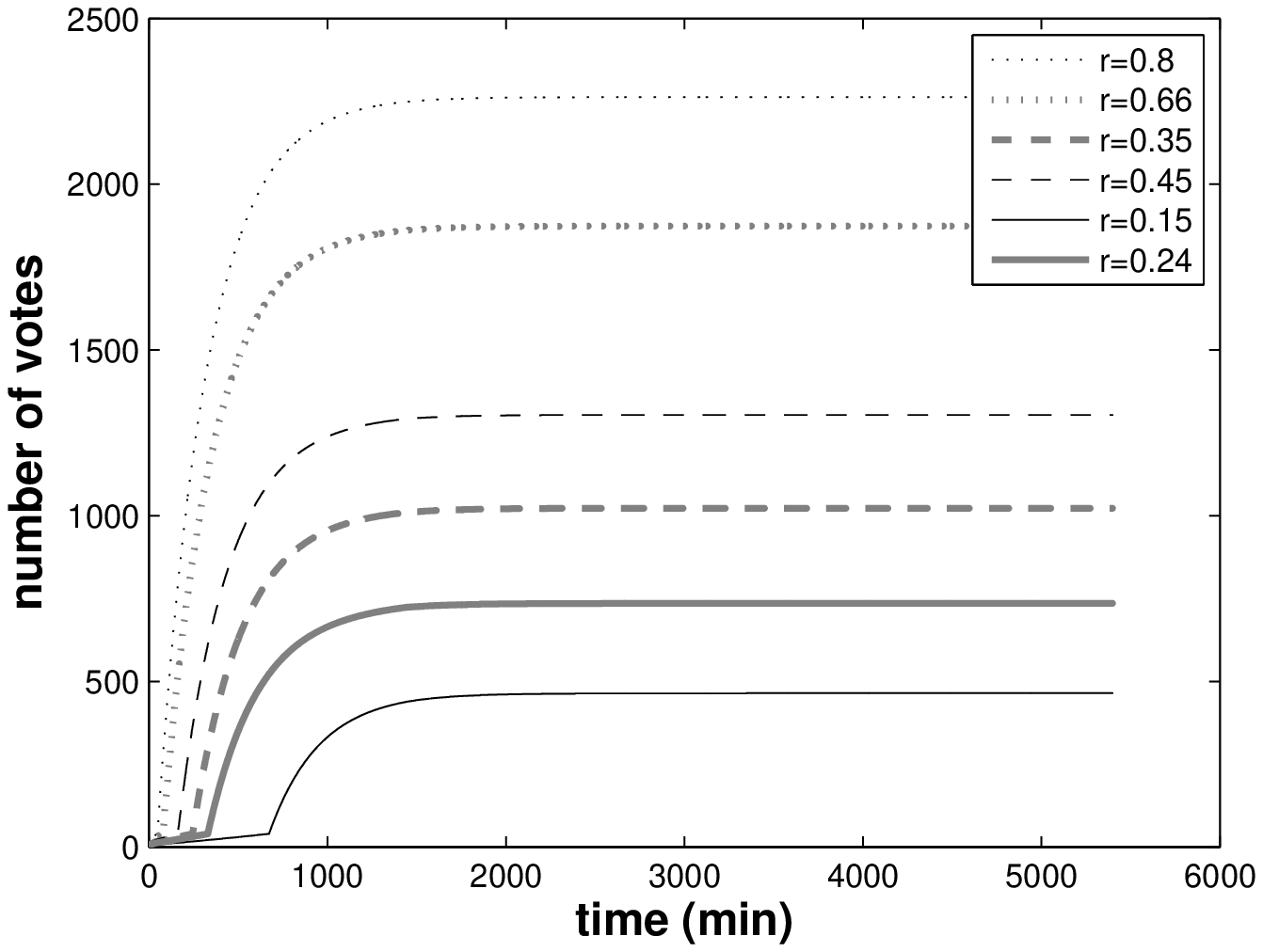} \\

  (a) & (b)
  \end{tabular}
\caption{(a) Evolution of the number of votes received by six
stories from the May dataset. The number of reverse friends the story
submitter has is given by $S$. (b) Predictions the model makes for
the same values of $S$ as (a). %(c) Dynamics of collaborative
%voting when variable threshold is used to promote stories to the front page.
} \label{fig:predictions}
\end{figure*}

Next, we consider the second modality of the Friends interface which
allows users to see the stories their friends voted on. This is the
situation described by the model Equation~\ref{eq:diggs}.
\figref{fig:predictions}(a) shows the evolution of the number of
votes received by six real stories from our dataset. $S$ denotes
the number of reverse friends the story's submitter had at the time
of submission. \figref{fig:predictions}(b) shows the model's
predictions for the same values of $S$ and different values of the
story interestingness parameter $r$. The addition of the new feature in
the Friends interface helps promote lower interest stories which
otherwise would not have been promoted. Overall there is qualitative
agreement between the data and the model, indicating that the basic
features of the Digg user interface we considered are enough to
explain the patterns of collaborative rating. The only significant
difference between the data and the model is visible in the bottom
two lines. In the data, the story posted by the user with $S=100$ is
promoted before the story posted by the user with $S=160$, but
saturates at smaller number of votes than the latter story. In the
model, the story with bigger $r$ is promoted first and reaches a
higher number of votes. The difference between data and the model is not surprising,
given the number of approximations made in the course of constructing the model
(see \secref{limitations} for discussion of modeling limitations).
For example, we assumed that the combined social network
of voters grows at the same rate for all stories. This cannot be true, obviously. If the
combined social network grew at a slower than assumed rate for the story posted
 by user with $S=160$, then this would explain the delay in being promoted to the front page.
Another effect which is not currently taken into consideration in the model is
that a story could have a different interestingness parameter $r$ to users
within submitter's social network than to
the general Digg audience. The model can be extended to include
inhomogeneous $r$.

\subsubsection{Modeling as a design tool} \label{sec:design}
Designing a complex system like Digg, which exploits the emergent behavior of many
independent evaluators, is exceedingly difficult. The choices made in the user interface,
for example, whether to allow users to see the stories their friends voted on or the most
popular stories within the last week or month, can have a dramatic impact on the
behavior of the collaborative rating system and on user experience. The designer has
to consider also the tradeoffs between story timeliness and interestingness,
how often stories are promoted, and the promotion algorithm to be
used. The promotion algorithm itself can have a dramatic impact on the
behavior of the collaborative rating system. As described in \secref{sec:promotion}, Digg's old promotion
algorithm alienated many users by making them feel that a cabal of top
users controlled the front page. Changes to the promotion algorithm in November 2006 appeared
to alleviate some of these concerns (while perhaps creating new ones).
Unfortunately, there are few tools, short of running the system,
that allow Digg developers to explore the various
options for the promotion algorithm.

We believe that
mathematical modeling and analysis can be a valuable tool for
exploring the design space of collaborative rating algorithms, despite the limitations described in \secref{limitations}.
We saw above that a story with low $r$ posted by a well connected user will be promoted to the front page.
If it is desirable to prevent uninteresting stories from getting to the front page, the
promotion algorithm could be changed to make it more difficult for people with bigger
social networks to get their stories promoted. Specifically, the promotion
threshold could be set to be a function of the the size of the submitter's social
network, e.g., the number of votes received by a story should be
$1.5S$ before the story is promoted.
In this case, stories with low $r$, e.g., $r=0.15$, will not be promoted to the front page even
when  they are posted by a well connected user with $S=160$.

\section{Dynamics of user rank}
\label{sec:rank}
 From its inception until February 2007 Digg ranked users according
to how successful they were in getting their stories promoted to the
front page. The more stories a user had on the front page, the
higher was his standing ($rank=1$ being the highest). If two users
had an equal number of front page stories, the one who was more
active (commented and voted on more stories) had higher rank. The
Top Users list was publicly available and offered prestige to those
who made it into the top tier. In fact, it is widely believed that
improving ones rank, or standing within the community, motivated
many Digg users to devote significant portions of their time to
submitting, commenting on and reading stories. Top users garnered
recognition as other users combed the Top Users list and made them
friends. They came to be seen as influential trend setters whose
opinions and votes were very valuable \cite{WSJ}. In fact, top users
became a target of marketers, who tried to pay them to promote their
products and services on Digg by submitting or voting on content
created by marketers. In an attempt to thwart this practice, in
February 2007 Digg discontinued making the Top Users list publicly
available.

\begin{figure}[tbh]
%\begin{tabular}{cc}
 \center \includegraphics[width=4in]{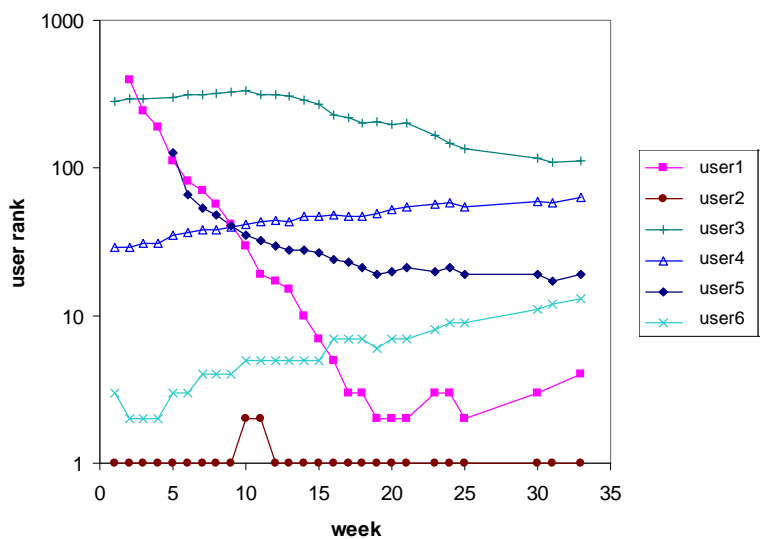}
%  & \includegraphics[height=1.9in]{rankvsfront} \\
%  (a) & (b)
%  \end{tabular}
\caption{ Evolution of user rank  } \label{fig:rank}
\end{figure}

We are interested in studying the dynamics of user
rank within the Digg community.  For our study we collected data about
the top ranked 1,020 Digg users weekly from May 2006 to January 2007.
For each user we extracted user's rank, the number of stories the
user submitted, commented and voted on, the number of stories that
were promoted to the front page, and the number of user's friends
and reverse friends (``people who have befriended the user''). We
reduced this data to 96 active users who made at least 50
submissions during any week. \figref{fig:rank} shows the change in
rank of six different users from the dataset. The top ranked user
($user2$) managed to hold on to that position for most of time, but
$user6$, who was ranked second at the beginning of the observation
period saw his rank slip to 10. Competition for top spots on the Top
Users list is very strong. Some users, such as $user1$ and $user5$,
came in with low rank but managed to reach the top tier of users by
week 25. Other users saw their rank stagnate ($user4$).

\subsection{Mathematical model of rank and social network dynamics}\label{sec:mathrank}
\begin{figure*}[tbh]
\center{
\begin{tabular}{cc}
  \includegraphics[width=2.0in]{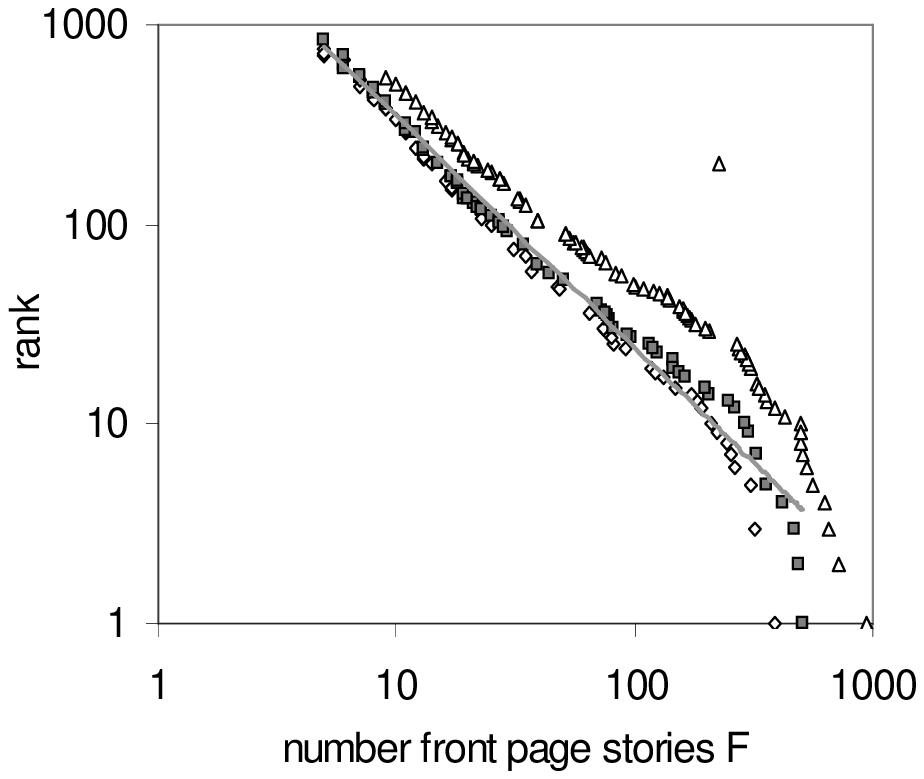} &
  \includegraphics[width=2.0in]{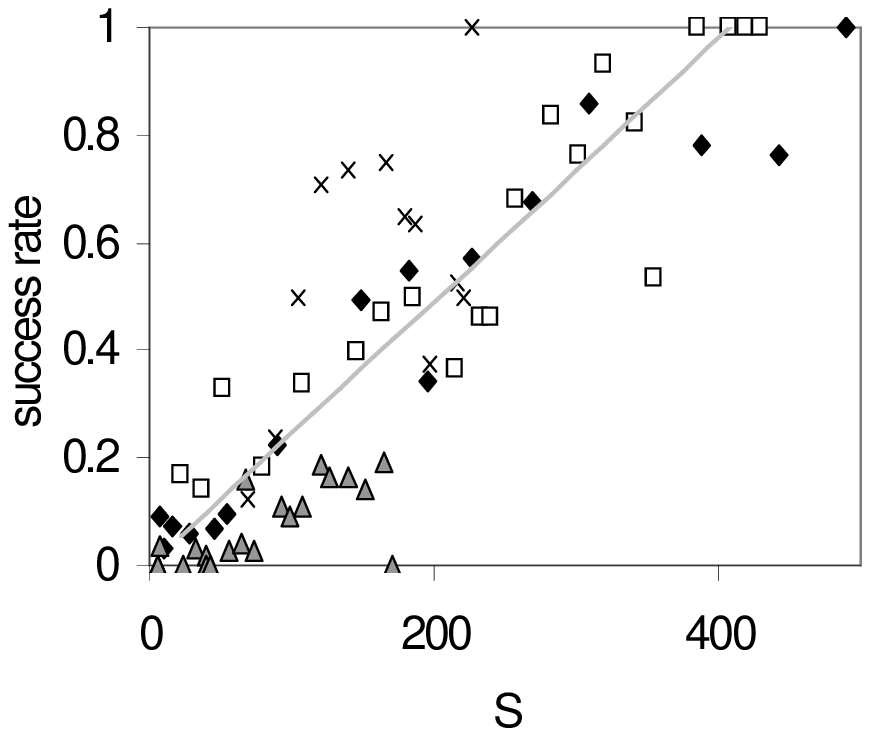} \\
(a) & (b) \\
  \includegraphics[width=2.0in]{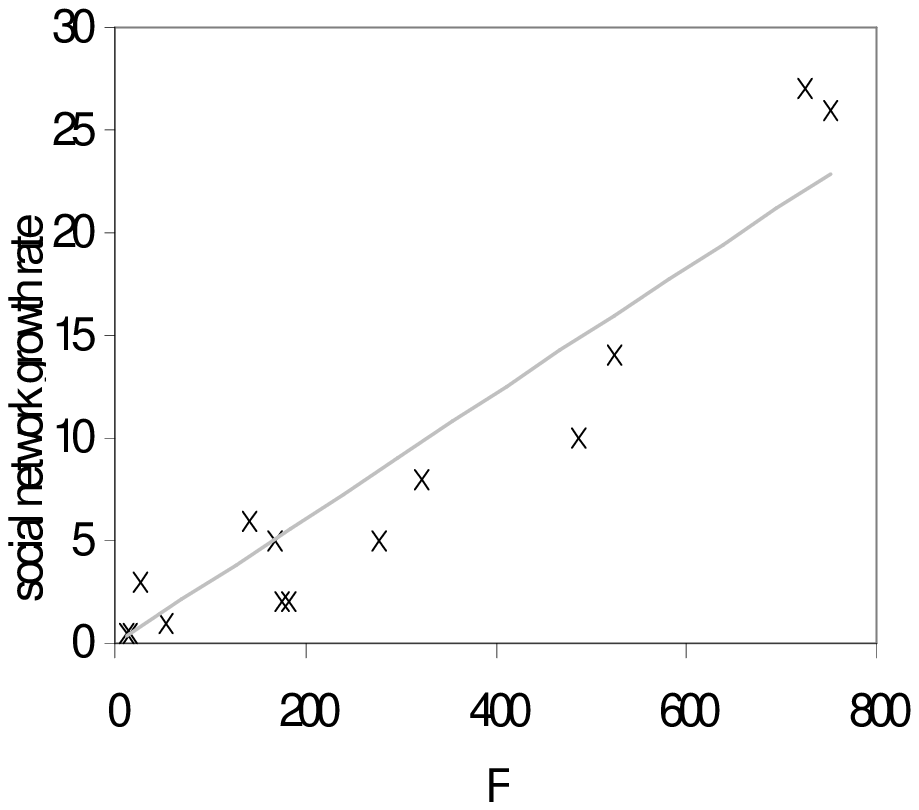} &
  \includegraphics[width=2.0in]{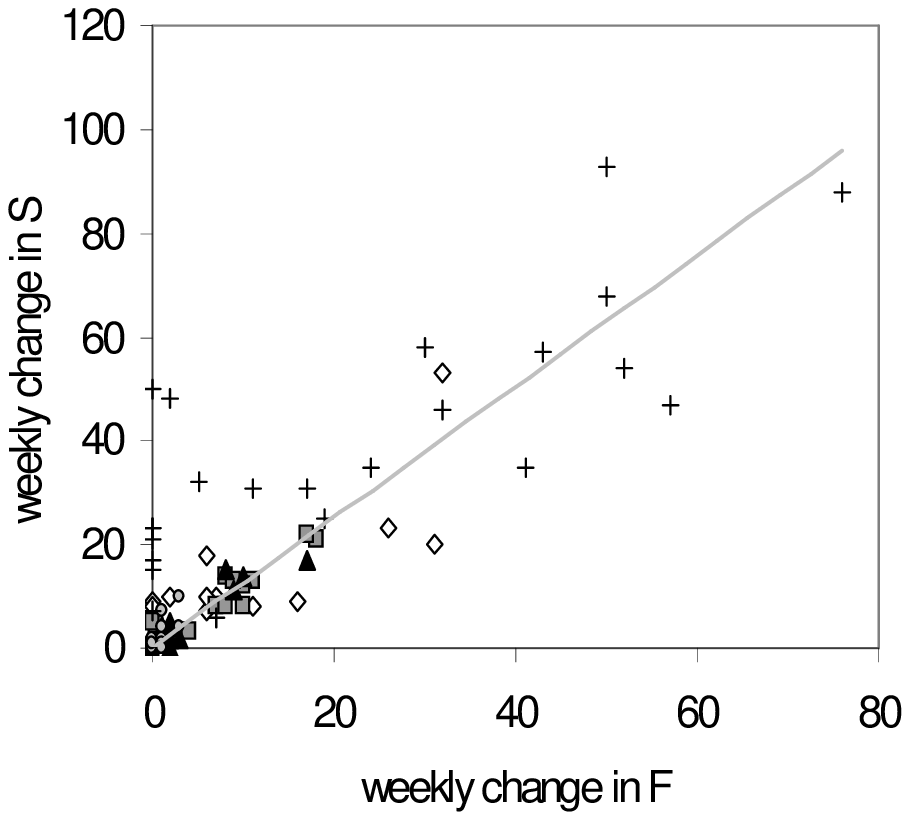} \\
     (c) & (d)
  \end{tabular}
  }
\caption{Parameter estimation from data. (a) Users' rank vs number
of their stories that have been promoted to the front page.
(b)Different users' success rates at getting their stories promoted
to the front page vs the number of reverse friends they have. In all
plot, solid lines represent fit to the data. (c) Temporal growth
rate of the number of user's reverse friends as a function of user
rank for the weeks when no new stories were submitted by these
users. (d) Weekly change in the size of the social network vs newly
promoted front page stories.  } \label{fig:sgrowth}
\end{figure*}

We are interested in creating a model that can predict how a user's
rank will change in time. As shown below, the model also describes
the evolutions of the user's personal social network, specifically,
how the number of reverse friends changes.
In addition to its explanatory power, the
model can be used to detect anomalies, for example, cases when a
user's rank, or social network, changes faster than expected due to collusion with other
users or other attempts to game the community. Because we do not know
the exact formula Digg uses to compute rank, we will use $F$, the number
of user's front page stories, as a
proxy for rank. \figref{fig:sgrowth}(a) plots user's rank vs the
number of front page stories for three randomly chosen users. The
data is explained well by a power law with exponent -1: i.e.,  $rank
\propto 1/F$.

The number of stories promoted to the front page clearly depends on
the number of stories a user submits, with the proportionality
factor based on the user's success rate. A user's success rate is
simply the fraction of the newly submitted stories that are promoted
to the front page. As we showed above, a user's success rate is
linearly correlated with the number of reverse friends he has ---
what we call social network size $S$. If $M$ is the rate of new
submissions made over a period of time $\Delta t=$week, then the
change in the number of new front page stories is
\begin{equation}
\label{eq:frontpage}
%\frac {dF}{dt}=c S(t) M
\Delta F(t)=c S(t) M \Delta t
\end{equation}
\noindent To estimate $c$, we plot user's success rate vs $S$ for
several different users, as shown in \figref{fig:sgrowth}(b).
Although there is scatter, a line with slope $c=0.002$ appears to
explain most of the trend in the data.

A given user's social network size $S$ is itself a dynamic variable,
whose growth depends on the rate other users discover him and add
him as a friend. The two major factors that influence a user's
visibility and hence growth of his social network are (i) his new
submissions that are promoted to the front page and (ii) his
position on the Top Users list. In addition, a user is visible
through the stories he submits to the upcoming stories queue and
through the commends he makes. We believe that these effects play a
secondary role to the two mentioned above. Therefore, the change in
the size of a user's social network can be expressed mathematically
in the following form:
\begin{equation}
\label{eq:Sgrowth}
%\frac {dS}{dt}=g(F)+ b \frac{dF}{dt}
%\Delta S(t)=a F(t) \Delta t + b \Delta F(t)
\Delta S(t)=g(F) \Delta t + b \Delta F(t)
\end{equation}
In order to measure $g(F)$, how a user's rank affects the growth of
his social network, we identified weeks during which some users made
no new submissions, and therefore, had no new stories appear on the
front page. In all cases, however, these users' social networks
continued to grow. \figref{fig:sgrowth}(c) plots the weekly growth
rate of $S$ vs $F$. There is an upward trend indicating that the
higher the user's rank (the greater the number of front page
stories) the faster his network grows. The grey line in
\figref{fig:sgrowth}(c) is a linear fit to the data of functional
form $g(F)=aF$ with $a=0.03$. \figref{eq:Sgrowth}(d) shows how newly
promoted stories affect the growth in the number of reverse contacts
for several users. Although there is variance, we take $b= 1.0$ from
the linear fit to the data.

\subsection{Solutions}\label{sec:ranksolutions}
\begin{figure*}[p]
\center{
\begin{tabular}{cc}
\multicolumn{2}{c}{data $\ \ \ \ \ \ \ \ \ \ \ \ \ \ \ $ user1 $\ \ \ \ \ \ \ \ \ \ \ \ \ \ \ $ model}\\
    \includegraphics[height=1.2in]{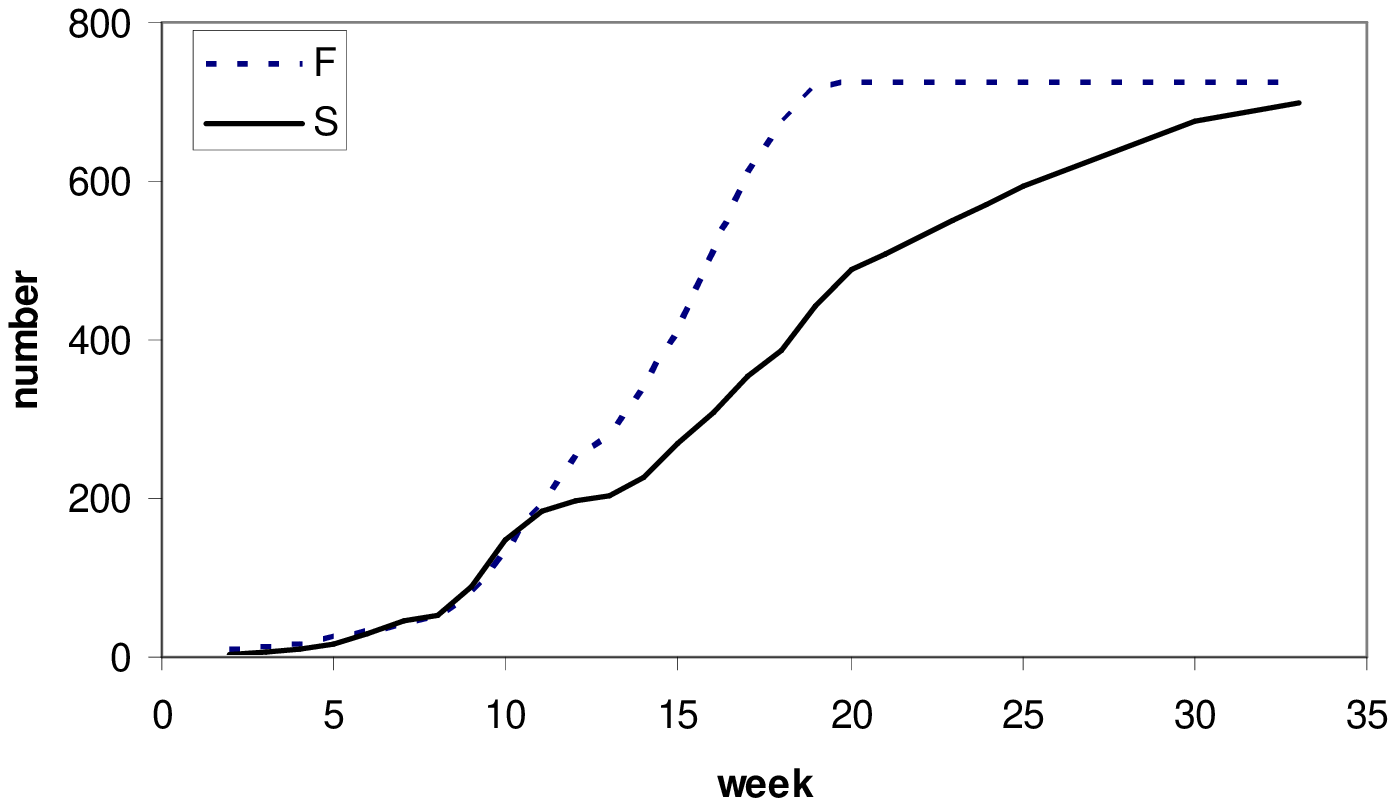} &
    \includegraphics[height=1.2in]{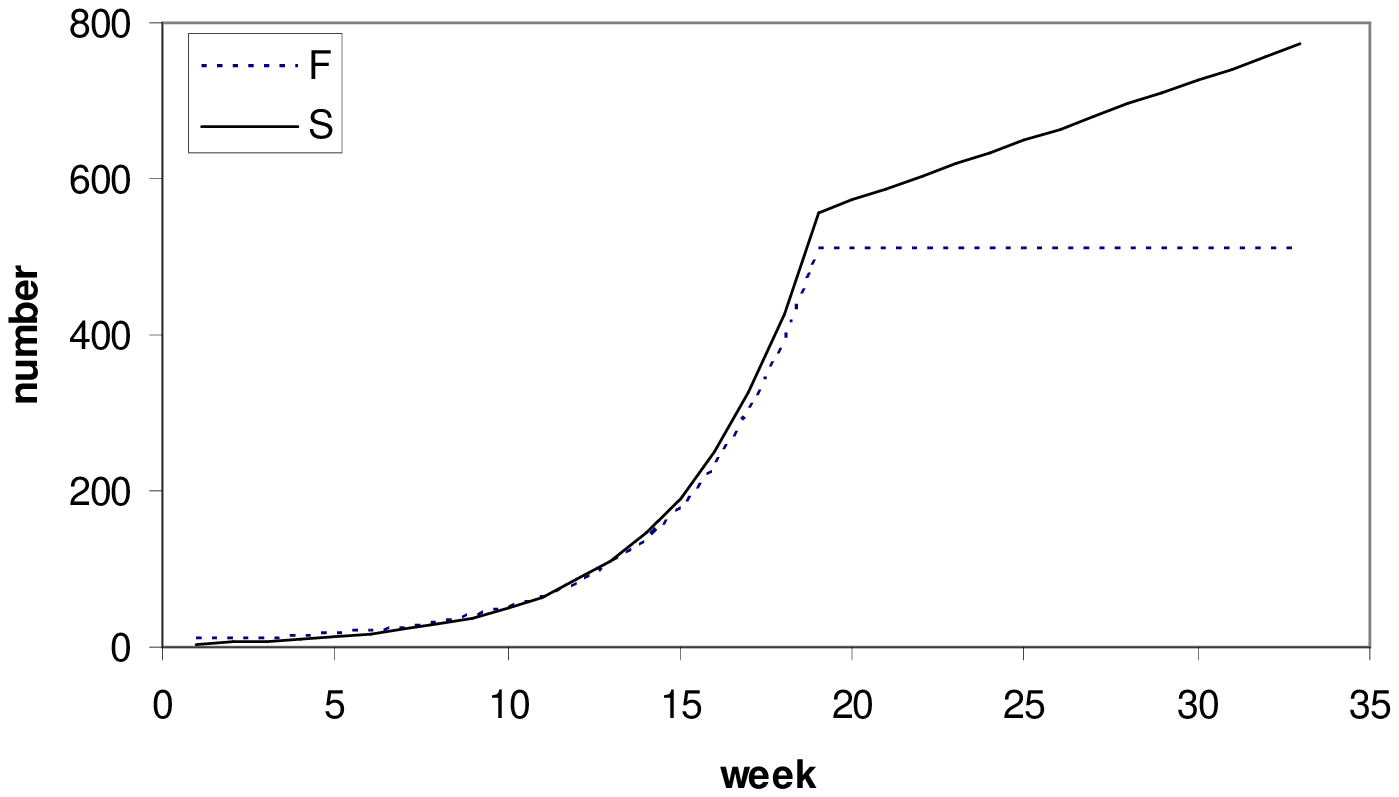} \\
\multicolumn{2}{c}{data $\ \ \ \ \ \ \ \ \ \ \ \ \ \ \ $ user2 $\ \ \ \ \ \ \ \ \ \ \ \ \ \ \ $ model}\\
    \includegraphics[height=1.2in]{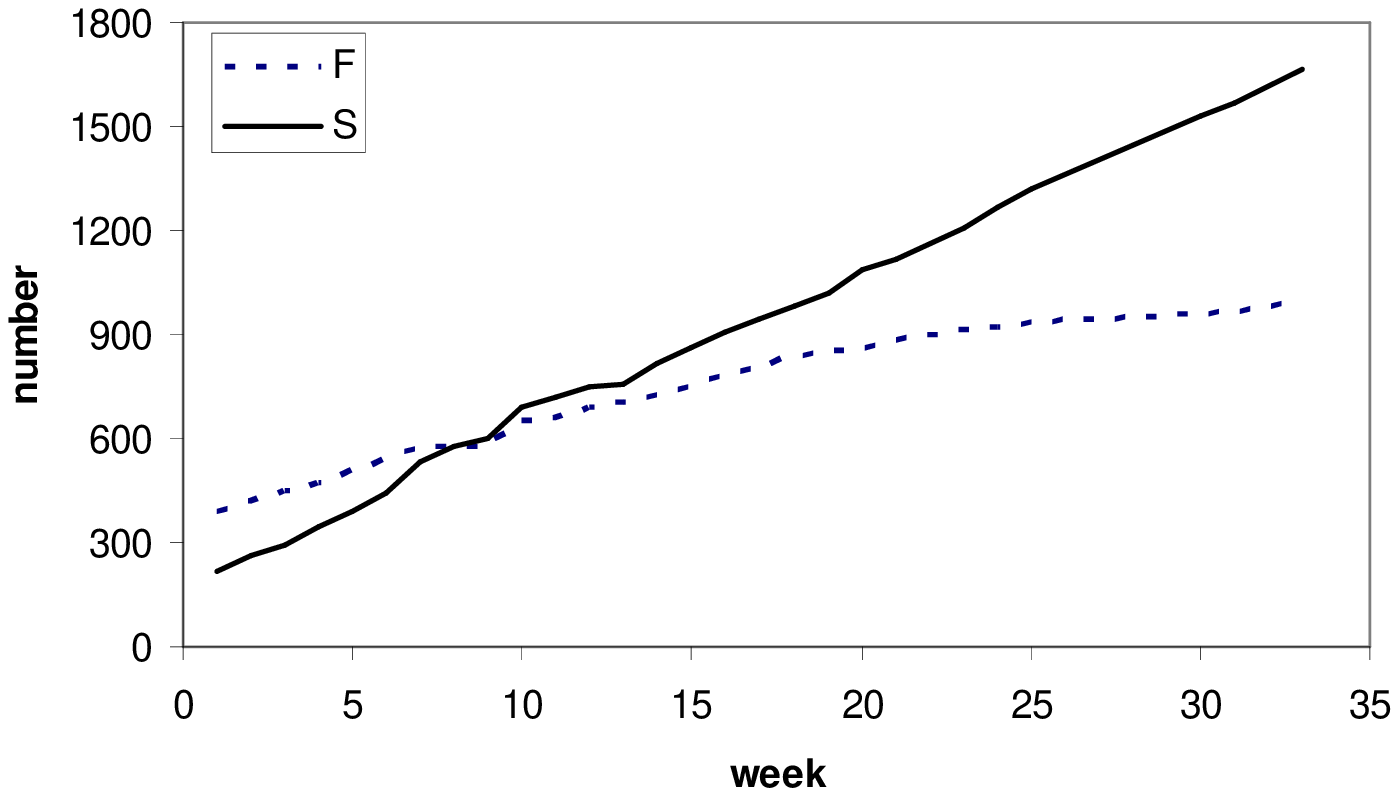} &
    \includegraphics[height=1.2in]{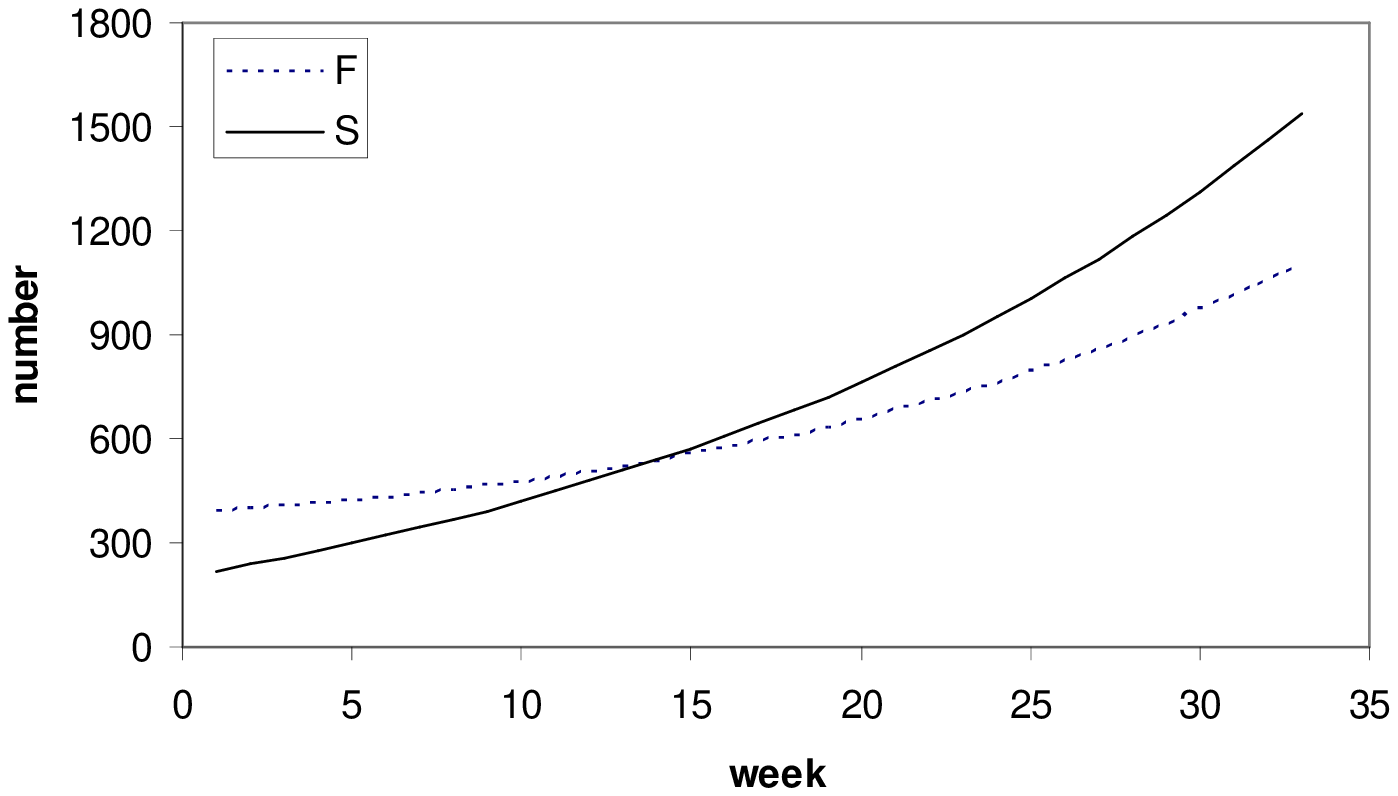} \\
\multicolumn{2}{c}{data $\ \ \ \ \ \ \ \ \ \ \ \ \ \ \ $ user3 $\ \ \ \ \ \ \ \ \ \ \ \ \ \ \ $ model}\\
    \includegraphics[height=1.2in]{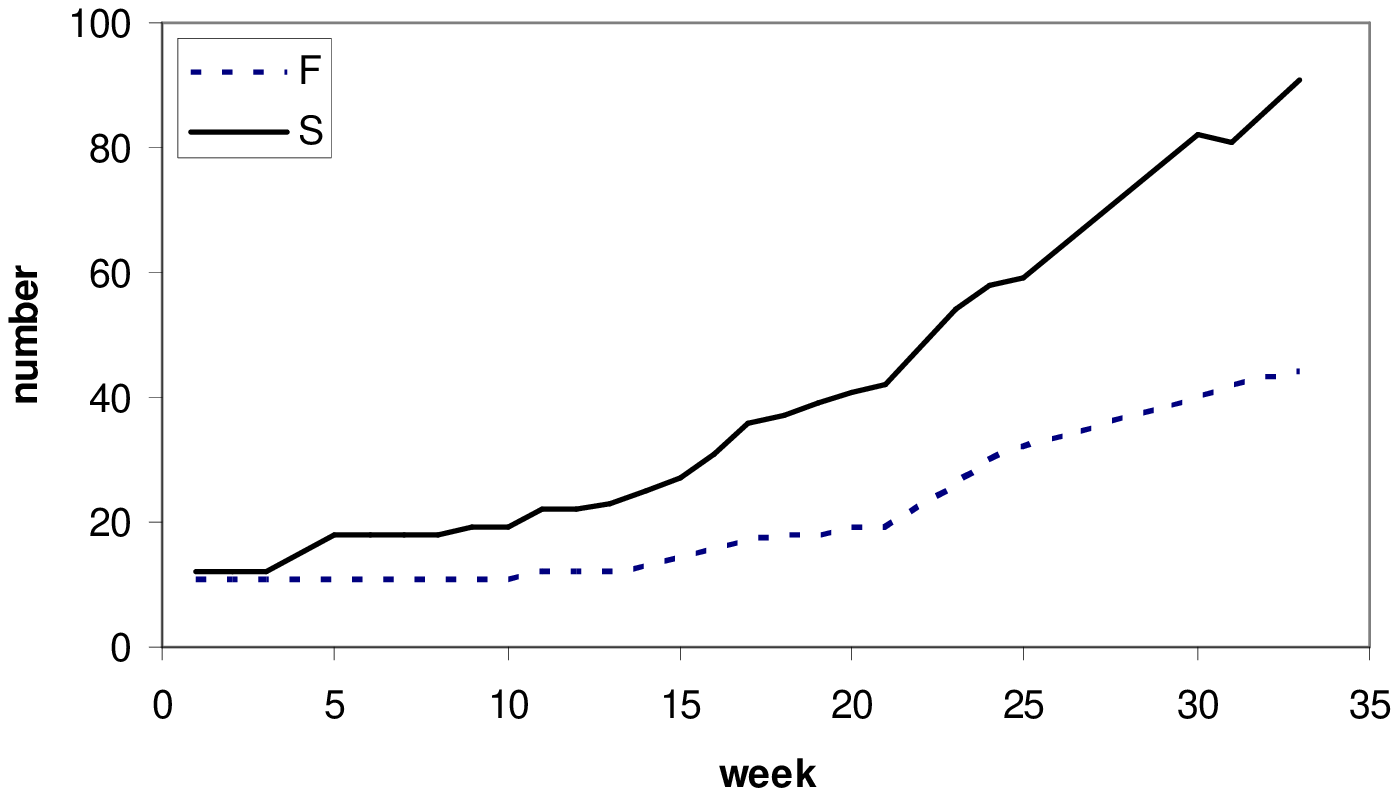} &
    \includegraphics[height=1.2in]{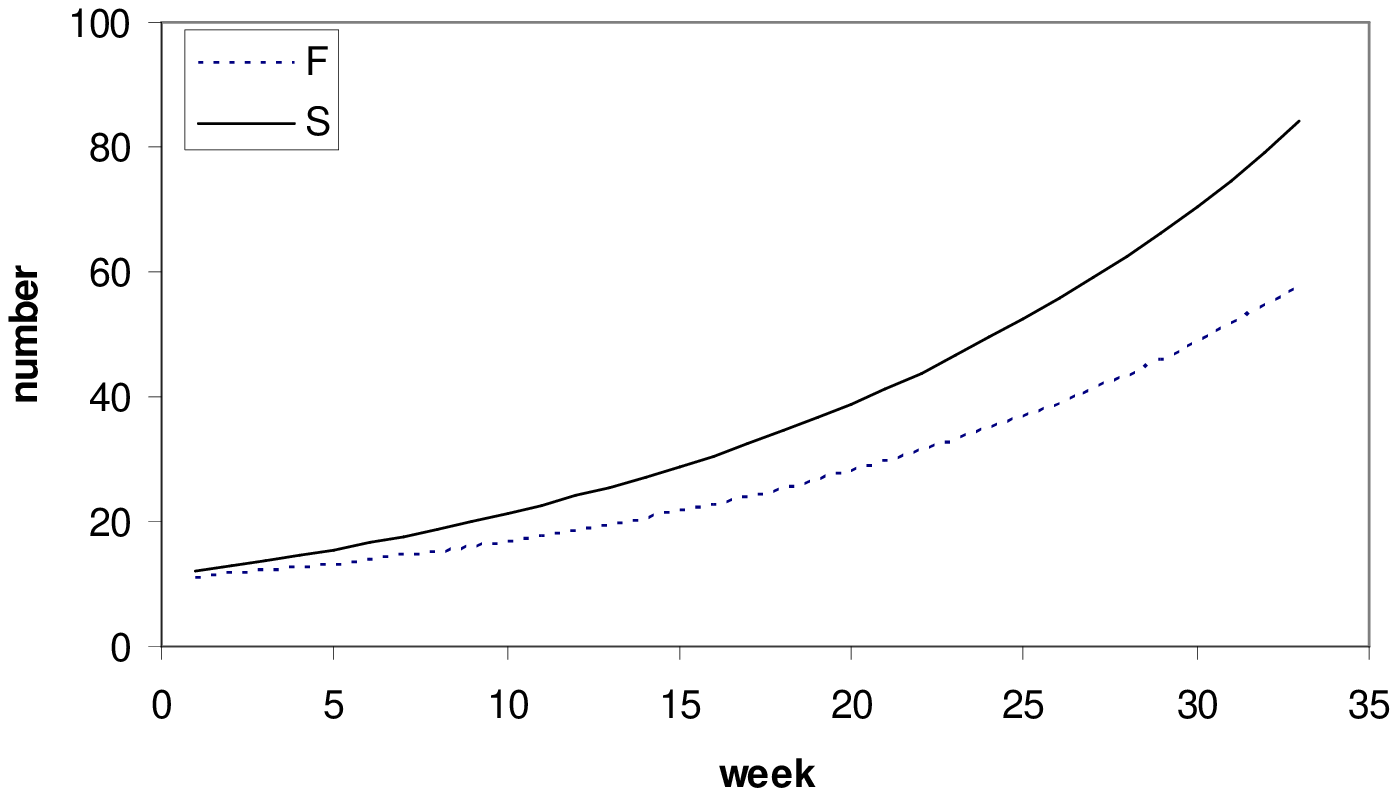} \\
\multicolumn{2}{c}{data $\ \ \ \ \ \ \ \ \ \ \ \ \ \ \ $ user4 $\ \ \ \ \ \ \ \ \ \ \ \ \ \ \ $ model}\\
    \includegraphics[height=1.2in]{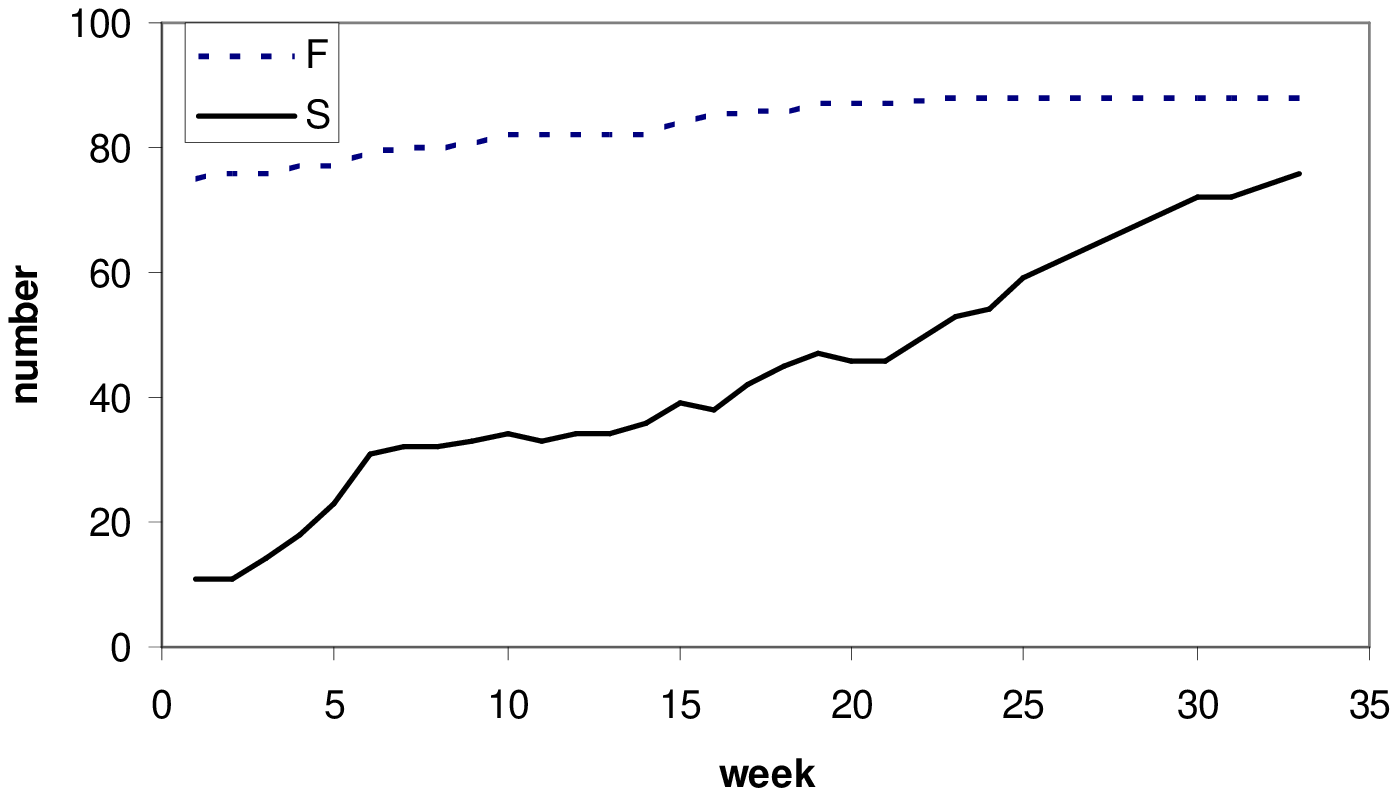} &
    \includegraphics[height=1.2in]{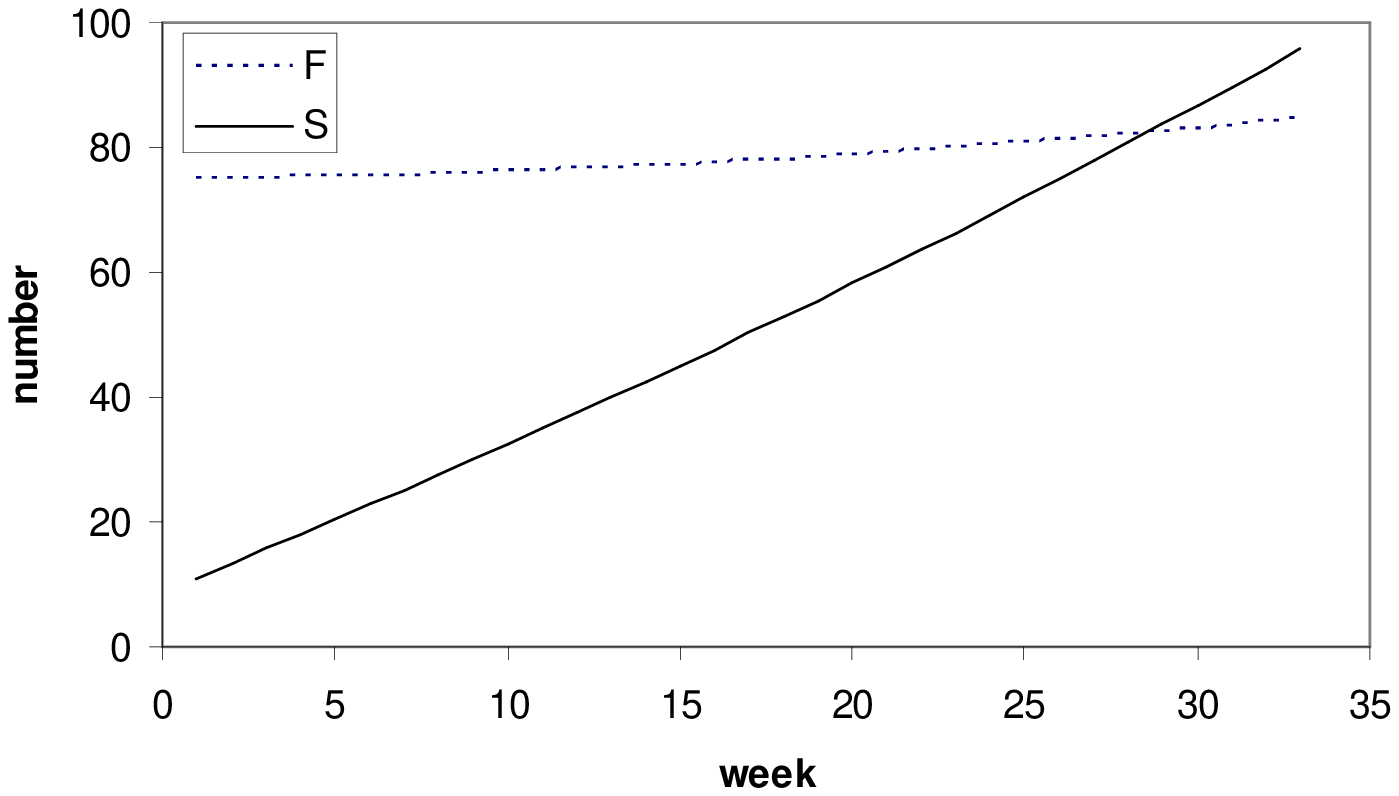} \\
\multicolumn{2}{c}{data $\ \ \ \ \ \ \ \ \ \ \ \ \ \ \ $ user5 $\ \ \ \ \ \ \ \ \ \ \ \ \ \ \ $ model}\\
    \includegraphics[height=1.2in]{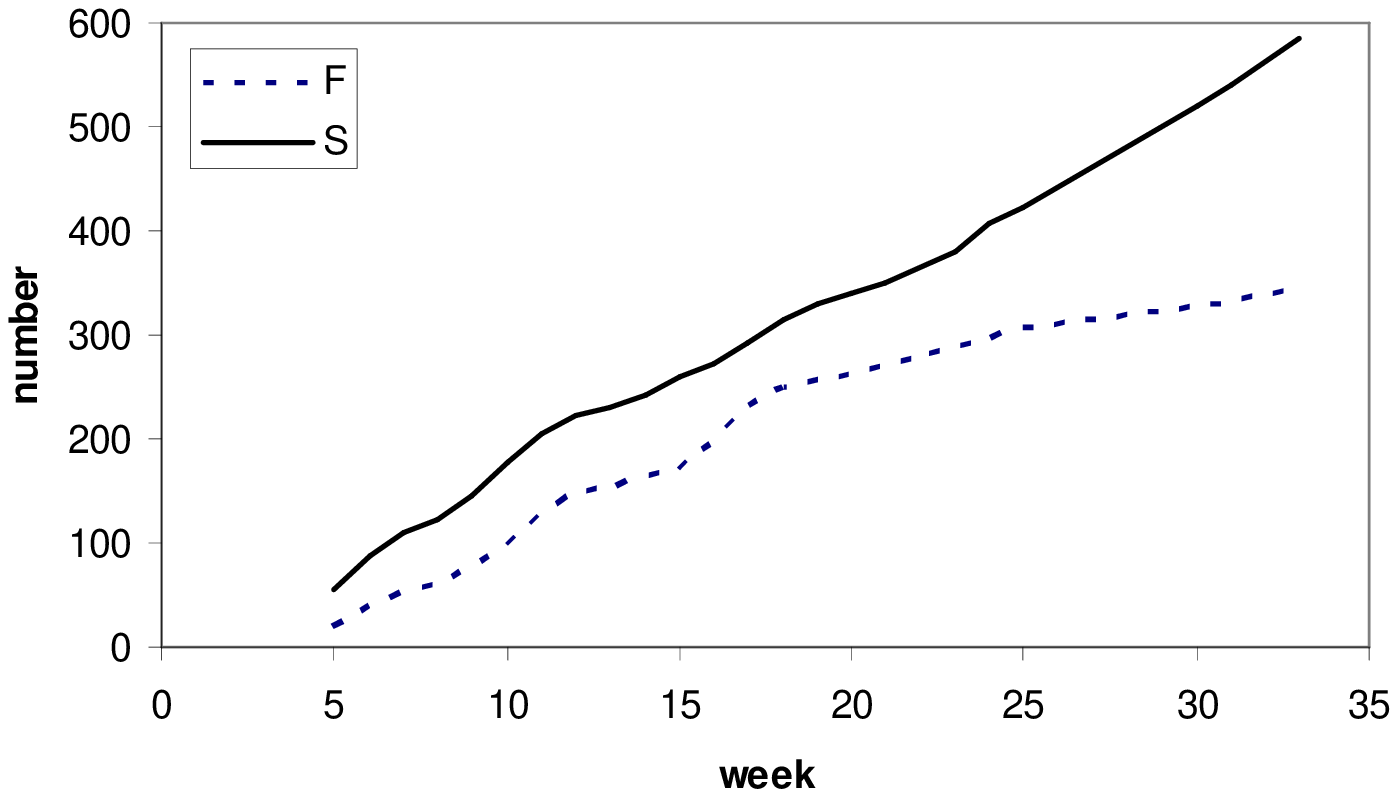} &
    \includegraphics[height=1.2in]{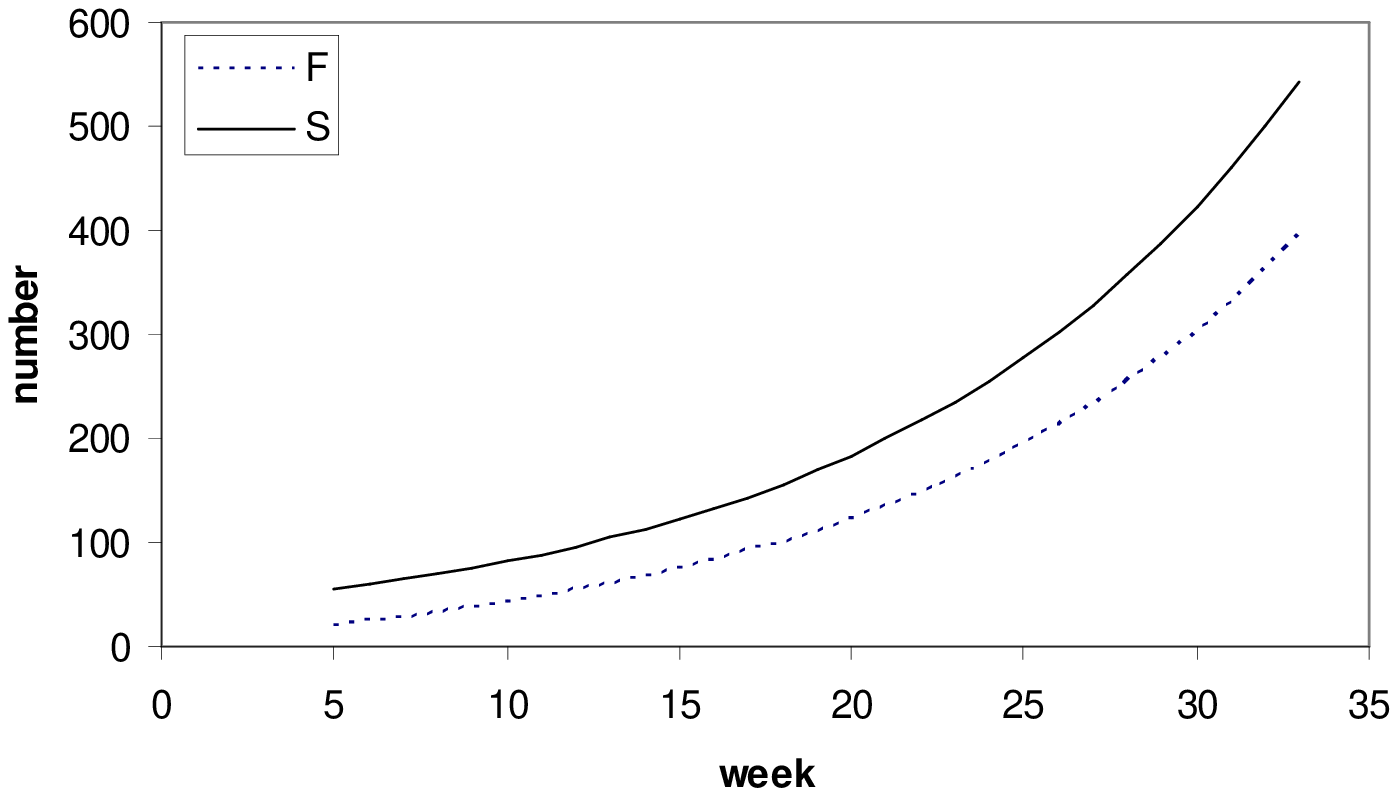} \\
\multicolumn{2}{c}{data $\ \ \ \ \ \ \ \ \ \ \ \ \ \ \ $ user6 $\ \ \ \ \ \ \ \ \ \ \ \ \ \ \ $ model}\\
    \includegraphics[height=1.2in]{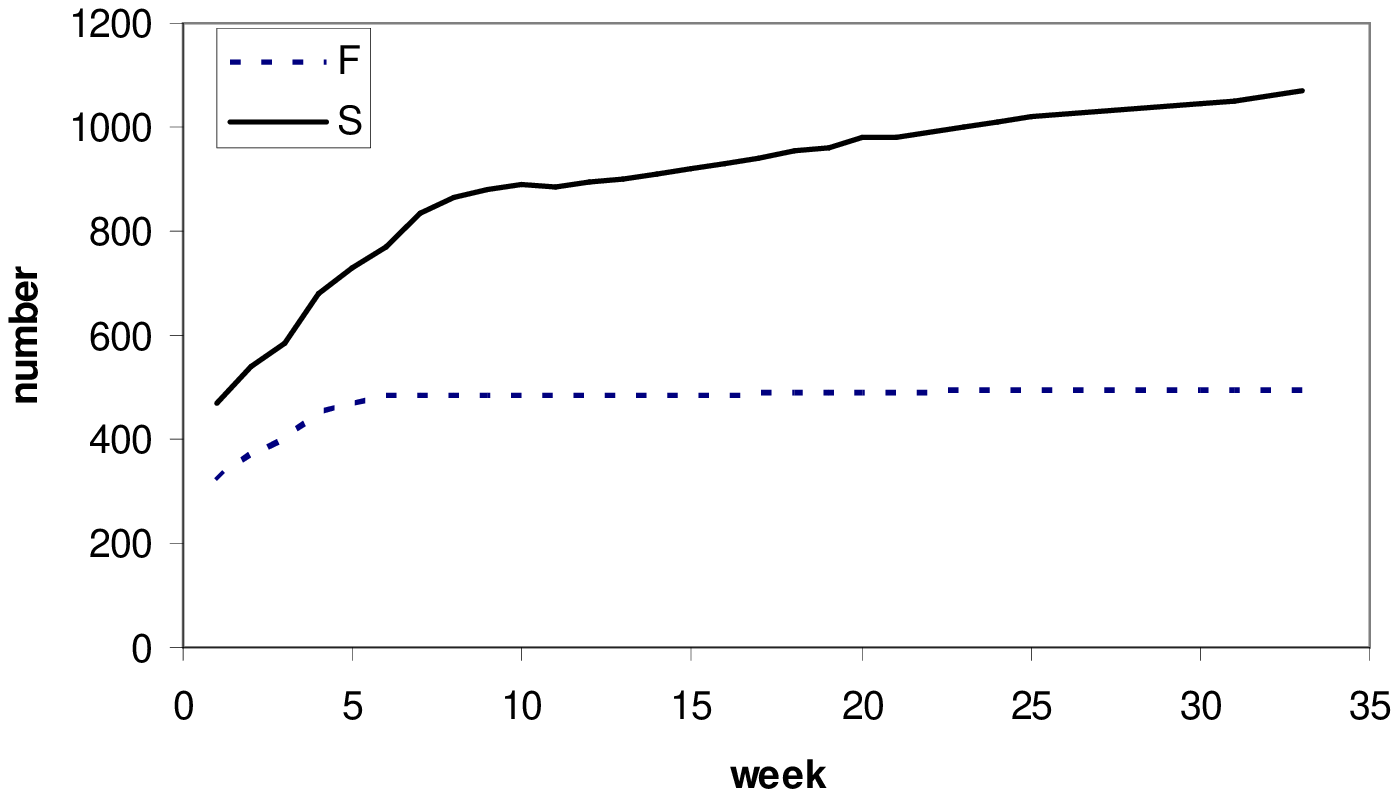} &
    \includegraphics[height=1.2in]{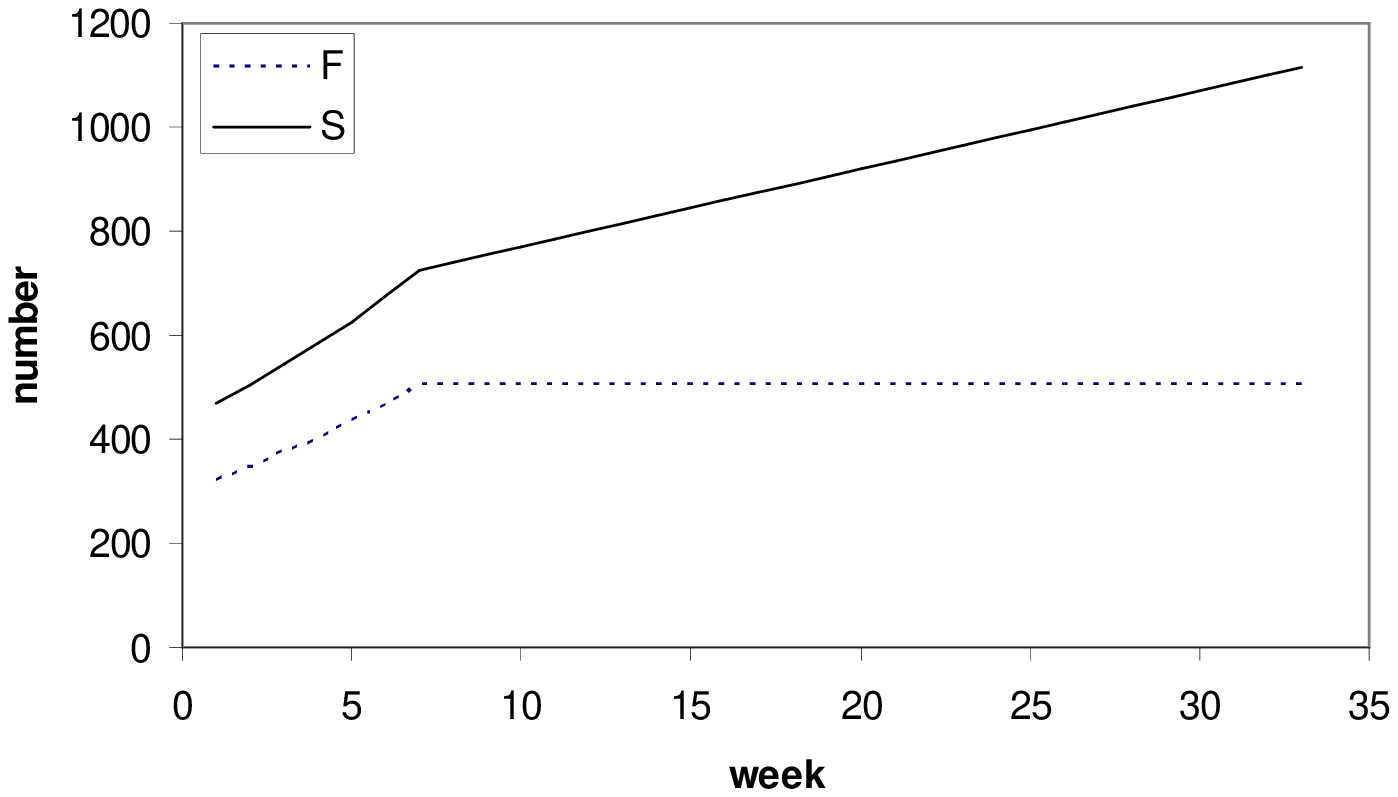}
  \end{tabular}}
\caption{Change over the course of 25 weeks of the number of front page stories and the number of
reverse friends a user has. The six users are the same ones shown in \protect{\figref{fig:rank}}.
The right hand plots show solutions to the rank dynamics model for each user. } \label{fig:rankgrowth}
\end{figure*}

%% after week 25, estimating number of reverse friends

\figref{fig:rankgrowth} shows how the personal social network (number of reverse friends) and
the number of front page stories submitted by six different users
from our dataset change in time. The users are the same ones whose
rank is shown in \figref{fig:rank}. To the right of each graph we
plot solutions to Equation~\ref{eq:frontpage} and Equation~\ref{eq:Sgrowth}. The
equations were solved under the initial conditions that $F$ and $S$
take the values they have at the beginning of the tracking period
for that user. We kept the submission rate $M$ fixed at its average weekly
value over the tracking period. The
actual submission rate fluctuates significantly for a given user from week
to week. We expect that including the actual submission rate will
substantially improve the agreement between the model and the data.

Solutions to the model qualitatively reproduce the important
features of the evolution of the user's rank and social network. Two
factors
--- user's activity via new submissions and the size of his social
network --- appear to  explain the change in user's rank. As long as
the user stays active and contributes stories to Digg, as
exemplified by $user2$, $user3$, $user4$ and $user5$, both the
number of promoted stories (rank) and the size of the user's social
network continue to grow. If a user stops contributing stories,
$user1$ and $user6$, his rank will stagnate as $F$ remains constant,
while his social network continues to grow, albeit at a slower rate.
Although a user can choose to submit more or fewer stories to Digg,
he cannot control the growth of his social network, e.g ., how and
when other users choose to make him a friend.\footnote{We suspect
that a user is able to influence the growth of his social network
through the implicit social etiquette of reciprocating friend
requests, but we have not yet been able to prove this conjecture.}
This helps promote independence of opinions, a key requirement of
the collaborative rating process, and raise the quality of ratings.
It appears, however, that the Top Users list serves to cement the
top tier position of the highest ranked users, since they continue
to grow their social networks, which in turn improves their success
rate. It will be
interesting to observe how elimination of the Top Users list alters
the Digg community and the quality of stories that appear on the
front page.

\section{Limitations of modeling} \label{limitations}

A number of assumptions and abstractions have been made in the
course of constructing the mathematical model and choosing its
parameters. Some of our assumptions affect the structure of the
model. For example, the only terms that contribute to the visibility
of the story come from users viewing the front page, upcoming
stories queue or seeing the stories one's friends have recently
submitted or voted on. There are other browsing modalities on Digg
that we did not include in the model. In the Technology section, for
example, a user can choose to see only the stories that received the
most votes during the preceding 24 hours (``Top 24 Hours'') or in
the past 7, 30 or 365 days. In the model, we only considered the
default ``Newly popular'' browsing option, which shows the stories
in the order they have been promoted to the front page. We assume
that most users choose this option. If data shows that other
browsing options are popular, these terms can be included in the
model to explain the observed behavior. Likewise, in the Friends
interface, a user can also see the stories his friends have
commented on or that are still in the upcoming queue, as well as the
stories they have submitted or voted on. We chose to include only
the latter two options from the Friends interface in our model.

In addition to the model structure, we made a number of assumptions
about the form of the terms and the parameters. The first model
describes the dynamics of votes an \emph{average} story receives. In
other words, it does not describe how the rating of a specific story
changes in time, but the votes on many similar stories averaged
together. Another point to keep in mind is that although there must
exist a large variance in Digg user behavior, we chose to represent
these behaviors by single valued parameters, not distributions.
Thus, we assume a constant rate users visit Digg, characterized by
the parameter $N$ in the model. We also assume that a story's
interestingness is the same for all users. In the model for rank
dynamics, all parameters were characterized by single value ---
taken to be the mean or characteristic value of the distribution of
user behavior. In future work we intend to explore how using
distributions of parameter values to describe the variance of user
behavior affects the dynamics of collaborative rating.

The assumptions we make help keep the models tractable, although a
question remains whether any important factors have been abstracted
away so as to invalidate the results of the model. We claim that the
simple models we present in the paper do include the most salient
features of the Digg users' behavior. We showed that the models
qualitatively explain some features of the observed collective
voting patterns. If we need to quantitatively reproduce experimental
data, or see a significant disagreement between the data and
predictions of the model, we will need to include all browsing
modalities and variance in user behavior. We plan to address these
issues in future research.

\section{Previous research}
Researchers have begun to study some aspects of social media. The most mature of
these research areas is the study of the blogosphere, e.g., as a means for
detecting trends in public opinion \cite{Adamic04}.  Tagging while still
new, is already attracting the interest of the scientific
community \cite{Golder05,boyd06}. While the initial purpose of tagging
was to help users organize and manage their own documents, it has
since been proposed that collectively tagging common documents can be
used to organize information via an informal classification system
dubbed a ``folksonomy'' \cite{Mika05}. The focus of our research, on the other hand,
is on the role of social networks in social information processing.

Many Web sites that provide information (or sell products or
services) use collaborative filtering technology to suggest relevant
documents (or products and services) to its users. Amazon and
Netflix, for example, use collaborative filtering to recommend new
books or movies to its users. Collaborative filtering-based
recommendation systems \cite{Konstan97grouplens} try to find users
with similar interests by asking them to rate products and then
compare ratings to find users with similar opinions. Researchers in
the past have recognized that social networks present in the user
base of the recommender system can be induced from the explicit and
implicit declarations of user interest, and that these social
networks can in turn be used to make new
recommendations \cite{ReferralWeb,perugini04}. Social media sites,
such as Digg, are to the best of our knowledge the first systems to
allow users to explicitly construct social networks and use them for
getting personalized recommendations. Unlike collaborative filtering
research, the topic of this paper was not recommendation per se, but
how social network-based recommendation affects the global rating of
information.

Social navigation, a concept closely linked to collaborative
filtering, help users evaluate the quality of information, or guide
them to new information sources, by exposing activities of other
users. Social navigation works ``through information traces left by
previous users for current users'' \cite{dieberger00} much like
footprints in the snow help guide pedestrians thro\-ugh a featureless
snowy terrain and pheromone trails left by ants help guide them to
food sources. Unlike our research, social navigation focused more on
information discovery rather than collaborative rating. Also,
despite strong analogy to pheromone-based navigation, no
mathematical analysis of social navigation has been done.

%math analysis
This paper borrows techniques from mathematical analysis of
collective behavior of multi-agent systems. Our earlier work
proposed a formal framework for creating mathematical models of
collective behavior in groups of multi-agent
systems \cite{Lerman04sab}. This framework was successfully applied
to study collective behavior in groups of
robots \cite{Lerman02,Martinoli04,Lerman06ijrr}.
Although the behavior of humans is, in general, far more complex than the behavior of robots, within the context of
a collaborative rating system, Digg users show simple behaviors that can be analyzed mathematically.
By comparing results of analysis with real world data extracted from Digg, we showed that mathematical
modeling is a feasible approach to study collective behavior of online users.

\section{Conclusion}

The new social media sites offer a glimpse into the future of the
Web, where, rather than passively consuming information, users will
actively participate in creating, evaluating, and disseminating
information. One novel feature of these sites is that they allow users to
create personal social networks of friends so as to easily keep track of friends' activities. Another novel feature is the
collaborative evaluation of content, either explicitly through voting or implicitly through
user activity. Together, these innovations lead to a new paradigm for interacting with information,
what we call \emph{social information processing}. Social information processing enables
users to collectively solve such hard information processing problems, such as document recommendation and filtering, and
evaluating the quality of documents.

We studied social information processing on the social news aggregator Digg. We showed that personal social networks
form the basis for an effective \emph{social recommendation} system, suggesting to users the stories his friends
have found interesting. Next, we studied collaborative voting of news stories on Digg, focusing
on the process by which the front page emerges by consensus from the distributed opinions of many voters.
We created a mathematical
model of the dynamics of collective voting and found that solutions of the model
qualitatively agreed with the evolution of votes received by actual stories on Digg.
We also studied how user's rank, which measures the influence of the user within the community,
changes in time as the user submits new stories and grows his social network. Again we found
qualitative agreement between data and model predictions.

Besides offering a qualitative explanation of user behavior,
mathematical modeling can be used as a tool to explore the design
space of user interfaces. The design of
complex systems such as Digg that exploit emergent behavior of large
numbers of users is notoriously difficult, and mathematical modeling
can help to explore the design space. It can help designers
investigate global consequences of different story promotion
algorithms before they are implemented. Should the promotion
algorithm depend on a constant threshold or should the threshold be
different for every story? Should it take into account the story's
timeliness or the popularity of the submitter, etc.?

Social media sites, such as Digg, show that it is possible to exploit the activities of others
to solve hard information processing problems. We expect progress in this field to
continue to bring novel solutions to problems in information processing, personalization,
search and discovery.

% Adapt2 IIS-0535182
% AutoSyn IIS-0413321
% Crowds BCS-0527725
\section*{Acknowledgements} This research is based on work supported
in part by the National Science Foundation under Award Nos.
IIS-0535182, IIS-0413321 and BCS-0527725. We are grateful to Fetch Technologies for providing
wrapper building and execution tools and to Dipsy Kapoor for assistance with data processing.

\bibliographystyle{plain}
\bibliography{../../social,../../lerman,../../robots}

\end{document}